\documentclass[12pt]{article}

\usepackage{latexsym}
\usepackage{verbatim}
\usepackage{amssymb}

\catcode`\@=11
\def\marginnote#1{}
\newcount\hour
\newcount\minute
\newtoks\amorpm
\hour=\time\divide\hour by60
\minute=\time{\multiply\hour by60 \global\advance\minute by-\hour}
\edef\standardtime{{\ifnum\hour<12 \global\amorpm={am}%
        \else\global\amorpm={pm}\advance\hour by-12 \fi
        \ifnum\hour=0 \hour=12 \fi
        \number\hour:\ifnum\minute<10 0\fi\number\minute\the\amorpm}}
\edef\militarytime{\number\hour:\ifnum\minute<10 0\fi\number\minute}
\def\draftlabel#1{{\@bsphack\if@filesw {\let\thepage\relax
   \xdef\@gtempa{\write\@auxout{\string
      \newlabel{#1}{{\@currentlabel}{\thepage}}}}}\@gtempa
   \if@nobreak \ifvmode\nobreak\fi\fi\fi\@esphack}
        \gdef\@eqnlabel{#1}}
\def\@eqnlabel{}
\def\@vacuum{}
\def\draftmarginnote#1{\marginpar{\raggedright\scriptsize\tt#1}}
\def\draft{\oddsidemargin -.5truein
        \def\@oddfoot{\sl preliminary draft \hfil
        \rm\thepage\hfil\sl\today\quad\militarytime}
        \let\@evenfoot\@oddfoot \overfullrule 3pt
        \let\label=\draftlabel
        \let\marginnote=\draftmarginnote
   \def\@eqnnum{(\theequation)\rlap{\kern\marginparsep\tt\@eqnlabel}%
\global\let\@eqnlabel\@vacuum}  }

\def\preprint{\twocolumn\sloppy\flushbottom\parindent 1em
        \leftmargini 2em\leftmarginv .5em\leftmarginvi .5em
        \oddsidemargin -.5in    \evensidemargin -.5in
        \columnsep 15mm \footheight 0pt
        \textwidth 250mmin      \topmargin  -.4in
        \headheight 12pt \topskip .4in
        \textheight 175mm
        \footskip 0pt
        \def\@oddhead{\thepage\hfil\addtocounter{page}{1}\thepage}
        \let\@evenhead\@oddhead \def\@oddfoot{} \def\@evenfoot{} }

\def\titlepage{\@restonecolfalse\if@twocolumn\@restonecoltrue\onecolumn
     \else \newpage \fi \thispagestyle{empty}\c@page\z@ 
        \def\thefootnote{\fnsymbol{footnote}} }

\def\endtitlepage{\if@restonecol\twocolumn \else  \fi
        \def\thefootnote{\arabic{footnote}}
        \setcounter{footnote}{0}}  

\catcode`@=12
\relax

\def\bea{\begin{array}}
\def\bem{\begin{displaymath}}
\def\beq{\begin{equation}}

\def\eea{\end{array}}
\def\eem{\end{displaymath}}
\def\eeq{\end{equation}}

\def\Im{\mathop{\rm Im}}

\def\ov{\overline}

\def\Re{\mathop{\rm Re}}

\def\s2w{\sin^2 \theta_W}

\relax
\def\dalpha{{\dot\alpha}}
\def\dbeta{{\dot\beta}}

\def\crbig{\\\noalign{\vspace {3mm}}}
\def\bigint{{\displaystyle\int}}

\def\Dint{\bigint d^2\theta d^2\ov\theta\,}
\def\Fint{\bigint d^2\theta\,}
\def\Fbarint{\bigint d^2\ov\theta\,}

\newcommand{\skipthispart}[1]{}

\relax

\begin{document}
\topmargin-2.1cm
\textwidth =450pt
\textheight 9.5in
%
%
%
%
\begin{titlepage}
\begin{flushright}
\today
\end{flushright}
\vspace{1.3cm}

\begin{center}{\Large\bf
Nonlinear \boldmath{${\cal N}=2$} Global Supersymmetry}

\vspace{1.3cm}

{\large\bf Ignatios Antoniadis $^{1,2}$\,\footnote{antoniad@lpthe.jussieu.fr}, Jean-Pierre Derendinger $^{1,2}$\,\footnote{derendinger@itp.unibe.ch} \\
and Chrysoula Markou $^2$\,\footnote{chrysoula@lpthe.jussieu.fr}  }

\vspace{6mm}

$^1$ Albert Einstein Center for Fundamental Physics \\
Institute for Theoretical Physics, University of Bern \\
Sidlerstrasse 5, CH--3012 Bern, Switzerland \\

\vspace{6mm}

$^2$ LPTHE, UMR CNRS 7589, Sorbonne Universit\' es,  \\ UPMC Paris 6, 75005 Paris, France

\end{center}
\vspace{1.3cm}

\begin{center}
{\large\bf Abstract}
\end{center}
\begin{quote}
We study the partial breaking of ${\cal N}=2$ global supersymmetry, using a novel formalism that allows for the off--shell nonlinear realization of the broken supersymmetry, extending previous results scattered in the literature. We focus on the Goldstone degrees of freedom of a massive ${\cal N}=1$ gravitino multiplet which are described by deformed ${\cal N}=2$ vector and single--tensor superfields satisfying nilpotent constraints. We derive the corresponding actions and study the interactions of the superfields involved, as well as constraints describing incomplete ${\cal N}=2$ matter multiplets of non--linear supersymmetry (vectors and single--tensors).

\end{quote}

\end{titlepage}
\renewcommand{\theequation}{\arabic{section}.\arabic{equation}}

\setcounter{footnote}{0}
\setcounter{page}{0}
\setlength{\baselineskip}{.6cm}
\setlength{\parskip}{.2cm}
\newpage
%
%

\tableofcontents








\section{Introduction} \label{secintro}
\setcounter{equation}{0}

The spontaneous breaking of global symmetries is described at low energies by a nonlinear $\sigma$--model of the corresponding Goldstone modes which have nonlinear transformations. These can often be obtained by applying an appropriate constraint on a linear $\sigma$--model. In the case of supersymmetry, the Goldstone modes are fermions, the goldstini, and the nonlinear $\sigma$--model for ${\cal N}=1$ is the Volkov--Akulov action~\cite{Volkov:1973ix}. In analogy with ordinary symmetries, it can be obtained (up to field redefinitions) by a chiral superfield $X$ satisfying a nilpotent constraint $X^2=0$ which eliminates its scalar component (sgoldstino) in terms of the goldstino bilinear~\cite{R, Lindstrom:1979kq, Casalbuoni:1988xh, komar}: 
\begin{equation}
X=-{\kappa \kappa \over 2F} +{\sqrt 2}\theta\kappa-\theta^2 F\, ,
\label{goldstinosf}
\end{equation}
where $\kappa$ is the two--component Goldstone fermion, $\theta$ the usual fermionic coordinates and $F$ the (nonzero) auxiliary field.
The most general K\"ahler potential is then quadratic $K=X{\ov X}$ and the superpotential linear in $X$, $P=\zeta X$, with a proportionality constant $\zeta$ fixing the scale of the supersymmetry breaking. Indeed, solving for $F$, one finds $F=\zeta + \textrm{fermions}$ and one obtains (on--shell) the Volkov--Akulov action~\cite{R,kuzenko}.


Besides the use of nonlinear supersymmetry as an effective low--energy theory at energies below the sgoldstino mass, it can also be realized exactly in particular vacua of type I string theory, when D--branes are combined with anti--orientifold planes that break the linear supersymmetries preserved by the D--branes, while they preserve the other half that are realized nonlinearly. In such vacua of ``brane supersymmetry breaking'', superpartners of brane excitations do not exist, and supersymmetry is nonlinearly realized with the presence of a massless goldstino in the open string spectrum~\cite{Antoniadis:1999xk, Kallosh:2015nia}.

The generalization of these results to extended supersymmetry, in particular to ${\cal N}=2$, broken at two different scales, is a challenging and not straightforward problem. An interesting case is ${\cal N}=2$ with one linear and one nonlinear supersymmetry, which is the standard situation of D--branes in a ${\cal N}=2$ supersymmetric bulk and describes the low--energy limit of partial ${\cal N}=2\to {\cal N}=1$ supersymmetry breaking. The goldstino of the nonlinear supersymmetry should then belong to a multiplet of the ${\cal N}=1$ linear supersymmetry, which can be either a vector or a chiral multiplet. In fact, both cases have to be studied, since they constitute the Goldstone degrees of freedom of a massive spin--3/2 multiplet. Indeed, a massless spin--3/2 multiplet contains a gravitino and a graviphoton, while a massive one contains, in addition, a spin--1 and a (Majorana) spinor, so that the Goldstone modes are a vector, two 2--component spinors and two scalars~\cite{Ferrara:1983gn}. 

When the second and nonlinear supersymmetry is taken into account, the above two ${\cal N}=1$ multiplets should be described by constrained ${\cal N}=2$ superfields associated with a Maxwell multiplet and a hypermultiplet. The latter comes with an extra complication since it has no off--shell formulation in the standard ${\cal N}=2$ superspace. Fortunately, the presence of bosonic shift symmetries associated with the would--be Goldstone bosons providing the longitudinal components of the spin--1 fields, implies that the chiral multiplet can be dualized to a linear multiplet having an off--shell description when promoted to a (constrained) ${\cal N}=2$ single--tensor superfield.

In this work we analyze the partial breaking of global ${\cal N}=2\to {\cal N}=1$ supersymmetry~\cite{APT}, extending known results in the literature on Maxwell multiplets~\cite{APT, BG, IZ, ADM, Kuz} and single--tensor multiplets~\cite{Bagger:1997pi, AADT}, we derive the corresponding ${\cal N}=2$ constrained superfields and study their possible interactions. The easiest way to introduce a breaking of ${\cal N}=2$ supersymmetry is by a (constant) deformation of the supersymmetry transformations of the fermions that cannot be absorbed in expectation values of the auxiliary fields, unlike the ${\cal N}=1$ case~\cite{ADM}. Partial breaking arises when the deformation parameters satisfy particular relations, guaranteeing the existence of one goldstino associated with a linear combination of the two supersymmetries. The goldstino superfield of one nonlinear supersymmetry can then be obtained by imposing a nilpotent (double chiral) constraint, in analogy with $X^2=0$ of ${\cal N}=1$. 

The outline of this paper is the following. In Section~2, we present a model of spontaneous partial breaking of ${\cal N}=2\to {\cal N}=1$ supersymmetry using one single--tensor multiplet, which contains a ${\cal N}=1$ linear multiplet $L$ and one chiral multiplet. The theory admits a special superpotential that allows for partial supersymmetry breaking, in analogy with the magnetic Fayet--Iliopoulos (FI) term in the Maxwell multiplet model of~\cite{APT}. This correspondence exchanges the ${\cal N}=1$ chiral field--strength superfield of the ${\cal N}=2$ Maxwell multiplet with the antichiral superfield $D_{\alpha}L$. Thus, the ${\cal N}=2$ Maxwell superfield is chiral under both supersymmetries (CC), while the single--tensor superfield is chiral under the first and antichiral under the second (CA). In Section~3, we discuss nonlinear deformations of the ${\cal N}=2$ Maxwell and single--tensor superfields, write the most general actions and compute the scalar potentials that have ${\cal N}=1$ supersymmetric minima. In Section~4, we consider the infinite--mass limit that freezes half of the degrees of freedom, and derive the constrained multiplets and the corresponding nilpotent constraints. We then give the solutions of the constraints (off--shell) and derive the generalizations of the goldstino Volkov--Akulov action in the presence of a linear supersymmetry, in addition to the nonlinear one. These are the supersymmetric Dirac--Born--Infeld (DBI) action and a similar action for the linear multiplet, in agreement with previous results. We then turn to the study of interactions. To this end, we introduce in Section~5 ``long" ${\cal N}=2$ superfields for the Maxwell and single--tensor multiplets with opposite relative chiralities compared to the ``short" ones, namely CA for the Maxwell and CC for the single--tensor, so that one can write a Chern--Simons type of interaction that we discuss in Section~6. This interaction leads to a super--Brout--Englert--Higgs mechansim without gravity, in which the linear multiplet is absorbed by the vector which becomes massive~\cite{AADT}. In Section~6, we also study more general constraints that describe incomplete ${\cal N}=2$ matter multiplets of non--linear supersymmetry (vectors or single--tensors), half of the components of which are projected out. Finally, Section~7 contains concluding remarks and open problems, while there are three appendices with our conventions (Appendix~A) and the technical details of the Maxwell multiplet (Appendices~B and C).

In the following, ${\cal W}$, ${\cal Z}$, ${\cal Y}\ldots$ denote ${\cal N}=2$ superfields with $8_B+8_F$ components, while hatted superfields $\widehat{\cal W}$, $\widehat{\cal Z}\ldots$ have $16_B+16_F$ fields. They are chiral with respect to the first supersymmetry (which shifts Grassmann coordinates $\theta_\alpha$) and either chiral or antichiral under the second supersymmetry (shifting $\widetilde\theta_\alpha$). All other superfields 
are ${\cal N}=1$ superfields. 

\section{Partial supersymmetry breaking with one hypermultiplet}
\setcounter{equation}{0}

In this Section we show the existence of partial supersymmetry breaking in a large class of ${\cal N}=2$ theories with a 
single hypermultiplet. The hypermultiplet couplings have a (translational) isometry allowing for a description in terms 
of a dual single--tensor multiplet which admits, like the Maxwell multiplet, a fully off--shell formulation. We use this
formulation to obtain these theories, dualize back to the hypermultiplet formulation and then display the strong 
similarity between partial breaking with a Maxwell (namely the APT model \cite{APT}) and partial breaking with a single--tensor multiplet.

The single--tensor ${\cal N}=2$ multiplet \cite{tensor1, LR, tensor2, tensor3} describes an antisymmetric tensor with gauge symmetry
\beq
\label{part1}
\delta\, B_{\mu\nu} = 2\, \partial_{[\mu}\Lambda_{\nu]},
\eeq
three real scalar fields and two Weyl (or massless Majorana) spinors. In the same manner that an antisymmetric tensor is dual to a pseudoscalar with 
axionic shift symmetry, a single--tensor multiplet is equivalent to a hypermultiplet with shift symmetry. In
both cases, the symmetry implies masslessness. In analogy with the Yang--Mills or Maxwell multiplet
but in contrast with the hypermultiplet, the single--tensor multiplet admits an off--shell formulation. 

In terms of ${\cal N}=1$ superfields, the single--tensor multiplet has two descriptions which can be viewed as 
the supersymmetrization either of the gauge invariant three--form field strength
\beq
\label{part2}
H_{\mu\nu\rho} = 3\,  \partial_{[\mu}B_{\nu\rho]}
\eeq
or of a two--form potential $B_{\mu\nu}$ and of its gauge transformation. The first description \cite{LR}
associates a real linear superfield $L$,
$\ov{DD}L=0$, which includes $H_{\mu\nu\rho}$, with a chiral superfield $\Phi$, $\ov D_\dalpha\Phi=0$, for a total
of $8_B+8_F$ off--shell fields. The second supersymmetry variations $\delta^*$ can be written as
\beq
\label{part3}
\delta^* L = -{i\over\sqrt2} (\eta D\Phi+\ov{\eta D}\ov\Phi)\,,
\quad\qquad
\delta^* \Phi = \sqrt2i\, \ov{\eta D}L\,,
\quad\qquad
\delta^* \ov\Phi = \sqrt2i\, \eta D L\,,
\eeq
where $\eta_\alpha$ is the spinor parameter of the second supersymmetry.
Since the linearity condition $\ov{DD}L=0$ is solved by
\beq
\label{part4}
L = D^\alpha\chi_\alpha - \ov D_\dalpha \ov\chi^\dalpha \equiv D\chi - \ov{D} \ov{\chi}\, ,
\eeq
where the chiral spinor superfield $\chi_\alpha$ includes $B_{\mu\nu}$, there is a second description with
two chiral superfields $\Phi$ and $Y$ associated with $\chi_\alpha$, for a total of $16_B+16_F$ 
fields.\footnote{The superfield $\Phi$ appears in both descriptions.} The variations are \cite{AADT}
\beq
\label{part5}
\begin{array}{rcl}
\delta^* Y &=& \sqrt2\, \eta\chi \, , 
\crbig
\delta^* \chi_\alpha &=& -{i\over\sqrt2} \Phi\,\eta_\alpha - {\sqrt2\over4} \eta_\alpha \, \ov{DD}\, \ov Y
-\sqrt2 i (\sigma^\mu\ov\eta)_\alpha \partial_\mu Y \, , 
\crbig
\delta^* \Phi &=&  2\sqrt2 i \left[\frac{1}{4}\,\ov{DD\eta\chi} 
+ i \partial_\mu\chi\sigma^\mu\ov\eta \right] .
\end{array}
\eeq
They close the ${\cal N}=2$ superalgebra off--shell. The supersymmetric extension of the gauge symmetry 
(\ref{part1}) is then
\beq
\label{part6}
\delta_{gauge} \chi_\alpha = -{i\over4} \ov{DD}D_\alpha \widehat V_2\,, \quad\qquad
\delta_{gauge} Y = {1\over2}\ov{DD}\, \widehat V_1\,, \quad\qquad
\delta_{gauge} \Phi = 0\,,
\eeq
with $\widehat V_1$ and $\widehat V_2$ real: the gauge transformation of the single--tensor multiplet in the description $(\chi_\alpha, \Phi, Y)$ is generated by a ${\cal N}=2$ Maxwell multiplet, which removes 
$8_B+8_F$ fields. There is a gauge with $Y=0$, residual ${\cal N}=1$ supersymmetry and gauge 
invariance generated by $\widehat V_2$.

The kinetic ${\cal N}=2$ lagrangian in the description $(L,\Phi)$ takes the simple form \cite{LR}
\beq
\label{part7}
{\cal L}_{kin.} = \Dint {\cal H}(L,\Phi,\ov\Phi) \,,
\eeq
where ${\cal H}$ is any real function solving the three--dimensional Laplace equation
\beq
\label{part8}
{\partial^2{\cal H}\over\partial L^2} + 2 {\partial^2{\cal H}\over\partial\Phi\partial\ov\Phi}=0.
\eeq
A unique superpotential $\widetilde{m}^2\Phi$ is allowed, since, under the second supersymmetry,
\beq
\label{part10}
\delta^* \Fint (\widetilde{m}^2\Phi) = \sqrt2i\,\widetilde{m}^2\Fint \ov{\eta D}\, L
\eeq
which is a derivative. For the real linear superfield $L$, $\ov D_\dalpha L$ is a chiral superfield with expansion
\beq
\label{part6b}
\ov D_\dalpha L = i\ov\varphi_\dalpha - (\theta\sigma^\mu)_\dalpha (v_\mu + i \partial_\mu C)
- \theta\theta (\partial_\mu\varphi\sigma^\mu)_\dalpha \,,
\qquad\qquad
v_\mu = {1\over6}\epsilon_{\mu\nu\rho\sigma}\,H^{\nu\rho\sigma}
\eeq
(in chiral coordinates), where the real scalar $C$ is the lowest component of $L$. Note also that the superpartner of $L$ (under the second supersymmetry) is
\beq
\label{isosw}
\Phi = z + \sqrt 2 \theta \psi - \theta \theta f\,.
\eeq

\subsection{Single-tensor multiplet formulation} \label{stfo}

To derive a theory with partial supersymmetry breaking, we first consider a generic ${\cal N}=1$ chiral function
$W(\Phi)$, with second supersymmetry variation
\beq
\label{part11}
\delta^*\Fint W(\Phi)= \sqrt2i\, \Fint W_\Phi\,\ov{\eta D}L\,, \qquad\qquad
W_\Phi = {dW\over d\Phi}\,.
\eeq
It is not a derivative unless $W(\Phi) \sim \Phi$.
Since\,\footnote{These equalities respect the first supersymmetry (which shifts $\theta$ and $\ov\theta$).}
\beq
\label{part12}
\ov{DD}\, (\ov{\theta\eta} \, L) = -2\, \ov{\eta D} \, L = \ov{DD}\, (\ov{\theta\eta} L + \theta\eta L) \,,
\eeq
the variation can also be written as\footnote{We usually omit derivatives when comparing lagrangian terms.}
\beq
\label{part13}
\delta^*\Fint W(\Phi) + {\rm h.c.} = 2\sqrt2i \Dint \Bigl[ W_\Phi  - \ov W_{\ov\Phi} \Bigr]
( \eta\theta + \ov{\eta\theta})L \,.
\eeq
Consider now the function 
\beq
\label{ini}
{\cal H}(L,\Phi,\ov\Phi)  = i \Bigl[ -L^2 [W_\Phi - \ov W_{\ov\Phi} ] + \ov\Phi W - \Phi\ov W\Bigr]\,,
\eeq
which is obviously a solution of the Laplace equation, while the action corresponding to
\beq
\label{part14}
\begin{array}{rcl}
{\cal L} &=& i\Dint\Bigl[ -L^2 [W_\Phi - \ov W_{\ov\Phi} ] + \ov\Phi W - \Phi\ov W\Bigr]+  \Fint ( \widetilde{m}^2\Phi) + {\rm h.c.} 
\crbig
&=& \displaystyle \Fint \Bigl[ {i\over2}W_\Phi\, (\ov DL)(\ov DL) 
- {i\over4} W \, \ov{DD}\,\ov\Phi + \widetilde{m}^2\Phi  \Bigr]
+ {\rm h.c.}
\end{array}
\eeq
is invariant under linear (off--shell) ${\cal N}=2$ supersymmetry.

To break spontaneously the second supersymmetry, we first add the generic superpotential $\widetilde{M}^2\, W(\Phi)$ to (\ref{part14}):
\beq
\label{part17}
\begin{array}{rcl}
{\cal L}_{nl} &=& \displaystyle \Fint \Bigl[ {i\over2}W_\Phi\, (\ov DL)(\ov DL) 
- {i\over4} W \, \ov{DD}\,\ov\Phi + \widetilde{m}^2\Phi +  \widetilde{M}^2 \, W \Bigr]
+ {\rm h.c.}
\crbig
&=& \displaystyle  \displaystyle 
i \Dint \Bigl[ -L^2 (W_\Phi-\ov W_{\ov\Phi}) + \ov\Phi W - \Phi\ov W \Bigr]
+  \Fint \Bigl[ \widetilde{m}^2\Phi +  \widetilde{M}^2 \, W \Bigr]
+ {\rm h.c.} 
\end{array}
\eeq
The action corresponding to (\ref{part17}) is then invariant under linear ${\cal N}=1$ supersymmetry as well as under the nonlinearly deformed second supersymmetry transformations
\beq
\label{part15}
\begin{array}{rcll}
\delta^*\, L &=& \displaystyle \delta^*_{nl} \, L - \frac{i}{\sqrt 2} (\eta D\Phi +\ov{\eta D}\ov \Phi), \qquad\qquad&
\delta^*_{nl} \, L \,\,=\,\, \sqrt2\,\widetilde{M}^2\,  (\ov{\theta\eta} + \theta\eta) ,
\crbig
&&& \delta^*_{nl} \, \ov D_\dalpha L \,\,=\,\, -  \sqrt2\, \widetilde{M}^2 \, \ov\eta_\dalpha \,,
\end{array}
\eeq 
with $\delta^*\Phi$ unchanged, since
\beq
\label{part16}
\delta^*_{nl}\, {\cal L}_{kin.} =  - i \sqrt2\,\widetilde{M}^2 \Fint W_\Phi\, \ov{\eta D}L + {\rm h.c.}
= - \widetilde{M}^2 \, \delta^* \Fint W(\Phi) + {\rm h.c.} 
\eeq
${\cal L}_{nl}$ depends on two complex numbers, the deformation parameter $\widetilde{M}^2$ and the quantity $\widetilde{m}^2$ in the linear ${\cal N}=2$ superpotential. Note also that the deformation in (\ref{part15}) implies that the spinor $\ov \varphi_\dalpha$ in the expansion (\ref{part6b}) of $\ov D_{\dot{\alpha}} L$ transforms like a goldstino. In fact, the transformations (\ref{part15}) for the ${\cal N}=1$ linear multiplet were first found in \cite{Bagger:1997pi} by performing a chirality switch on the transformations of the ${\cal N}=1$ Maxwell multiplet, first given in \cite{BG}. 

\subsubsection{Alternative proof}

Let us consider the ${\cal N}=2$ supersymmetric lagrangian (\ref{part7}). Suppose that, to induce the partial breaking, we deform the second supersymmetry transformations of the single--tensor multiplet, in such a way that the spinor $\ov \varphi_\dalpha$ in the expansion of $\ov D_{\dot{\alpha}} L$ transforms like a goldstino; the transformations take then the form (\ref{part15}). The deformation induces a new term in the variation of the lagrangian under the second supersymmetry:
\begin{equation} 
 \delta^*_{def} \mathcal{L}_{kin.} =   \sqrt 2 \, \widetilde{M}^2   \Dint  \, \mathcal{H}_L  (\theta \eta  + \ov{\theta} \ov{\eta} ) \, ,
\label{deff}
\end{equation}
where $\mathcal{H}_L=\frac{\partial \mathcal{H}}{\partial L}$ and $\mathcal{H}$ satisfies the Laplace equation in the limit $\widetilde{M}^2  \rightarrow 0$. 
The expression (\ref{deff}) selects the $\theta\theta\ov\theta$ and $\ov{\theta\theta}\theta$ components
of ${\cal H}_L$. To obtain partial breaking, these components must transform as derivatives under the first, unbroken supersymmetry. This is the case if the highest component of ${\cal H}_L$ is zero or a derivative, 
\begin{equation}
\int d^2  \theta d^2 \ov{\theta} \,\mathcal{H}_L  = {\rm derivative} \, ,
\end{equation}
whose solution is 
\beq
\mathcal{H}_L  = \widetilde{\mathcal{G}}(\Phi) + \ov{\widetilde{{\mathcal{G}}}}(\ov{\Phi}) - 2L \Big( {\mathcal{G}}_{\Phi \Phi}(\Phi) + \ov{{\mathcal{G}}}_{\ov{\Phi} \ov{\Phi}}(\ov{\Phi})\Big)
\eeq
where $\mathcal{G}$, $\widetilde{\mathcal{G}}$ are holomorphic functions of $\Phi$ and $\mathcal{G}_\Phi=\frac{d}{d\Phi}\mathcal{G}(\Phi)$ (we use the derivatives merely for convenience). The prefactor $-2$ of $L$ terms is conventional. Consequently, 
\begin{equation}
 \mathcal{H} = K(\Phi,\ov{\Phi}) + L \Big( \widetilde{\mathcal{G}}(\Phi) + \ov{\widetilde{{\mathcal{G}}}}(\ov{\Phi}) \Big) - L^2\Big( {\mathcal{G}}_{\Phi \Phi}(\Phi) + \ov{{\mathcal{G}}}_{\ov{\Phi} \ov{\Phi}}(\ov{\Phi})\Big)  \,,
\end{equation}
where $K(\Phi,\ov{\Phi})$ is a function of $\Phi$, $\ov{\Phi}$ and, using the Laplace equation, we obtain
\begin{equation}
 \mathcal{H} =   \Big(\ov{\Phi} \mathcal{G}_\Phi(\Phi) + \Phi \ov{\mathcal{G}}_{\ov{\Phi}}(\ov{\Phi}) \Big)- L^2 \Big( {\mathcal{G}}_{\Phi \Phi}(\Phi) + \ov{{\mathcal{G}}}_{\ov{\Phi} \ov{\Phi}}(\ov{\Phi})\Big) \, ,
\label{sol}
\end{equation}
since terms linear in $L$ do not contribute to the integral $ \Dint $.

Now let us consider again the derformation (\ref{deff}) of the lagrangian. With the use of (\ref{sol}), it becomes (since terms proportional to $L^0$ do not contribute):
\beq
\begin{array}{rcl}
{ \delta^*_{def} \mathcal{L}_{kin.}} &=& - 2 \sqrt 2 \widetilde{M}^2 \Dint \, L  \,\Big( {\mathcal{G}}_{\Phi \Phi}(\Phi) + \ov{{\mathcal{G}}}_{\ov{\Phi} \ov{\Phi}}(\ov{\Phi})\Big)  (\theta \eta  + \ov{\theta} \ov{\eta} )
\crbig
&=& 
  \frac{\widetilde{M}^2 }{\sqrt 2 } \Fint \, \ov{D} \ov{D} \Big[ L \, {\mathcal{G}}_{\Phi \Phi}(\Phi)  \,\ov{\theta} \ov{\eta}  \Big] + \textrm{h.c.}
\crbig
&=&  - \widetilde{M}^2 \sqrt 2  \Fint \, ( \ov{\eta} \ov{D}L) \,{\mathcal{G}}_{\Phi \Phi}(\Phi)  + \textrm{h.c.} =  i \widetilde{M}^2 \, \delta^* \Fint \, {\mathcal{G}}_\Phi(\Phi)  + \textrm{h.c.} 
\end{array}
\eeq
Consequently, the deformed lagrangian
\begin{equation}
\mathcal{L}_{def,kin.} = \Dint \, \mathcal{H} (L, \Phi, \ov{\Phi}) - i \widetilde{M}^2 \Fint \,{\mathcal{G}}_\Phi(\Phi)  + \textrm{h.c.} 
\end{equation}
is invariant under the first (linearly--realized) supersymmetry as well as under the second nonlinearly--realized one. It is also obvious that the lagrangians corresponding to (\ref{ini}) and (\ref{sol}) are equivalent upon identifying ${\cal G}_\Phi (\Phi) = i W(\Phi)$.

\subsubsection{The vacuum}

Theory (\ref{part17}) with $\widetilde{m}^2=0$ can be derived from a deformed chiral--antichiral ${\cal N}=2$ superfield with the use of a prepotential function ${\cal G}(\mathcal{Z})$.
Let us define\footnote{We introduce a second set of Grassmann coordinates $\widetilde{\theta}_{\alpha}, \ov{\widetilde\theta}_{\dot{\alpha}}$ and use chiral--antichiral coordinates $\widetilde{y}^\mu$ such that $\ov{D}_{\dot{\alpha}} \widetilde{y}^\mu = \widetilde{D}_\alpha \widetilde{y}^\mu =0$. Then, $\mathcal{Z}$ is a function of $\widetilde{y}^\mu$, $\theta$ , $\ov{\widetilde{\theta}}$.}
\beq
\label{part18}
{\cal Z} = \Phi + \sqrt2i \, \ov{\widetilde\theta D} L - {1\over4}\,\ov{\widetilde\theta\widetilde\theta}\, 
\Bigl[ 4 i \widetilde{M}^2 + \ov{DD}\,\ov\Phi\Bigr] \,.
\eeq
We then obtain 
\beq
\label{part19}
\Fint \bigint d^2\ov{\widetilde\theta}\, {\cal G}({\cal Z}) + {\rm h.c.} =
\Fint \Biggl[ {1\over2}\, {\cal G}_{\Phi\Phi} \, (\ov DL)(\ov DL)
-{1\over4}{\cal G}_\Phi\, \ov{DD}\, \ov\Phi 
-i \, \widetilde{M}^2 \, {\cal G}_\Phi
\Biggr] + {\rm h.c.}
\eeq
Clearly, ${\cal G}_\Phi (\Phi) = i W(\Phi)$. Notice that the deformation cannot be understood as the expectation value of 
a scalar of the ${\cal N}=1$ superfields.

Partial supersymmetry breaking is achieved if theory (\ref{part17}) has a vacuum state invariant under the first
(linear) supersymmetry. We then analyze the scalar potential, which, since $L$ does not have auxiliary fields, 
follows from the auxiliary $f$ (in $\Phi$) only. The auxiliary field lagrangian is\footnote{In this Section, we use the same 
notation $\Phi$ for the superfield and its lowest component. The other components are $\psi$ and $f$, as in the other Sections. The kinetic metric of the multiplet is $i \big( W_\Phi - \ov W_{\ov \Phi}\big)$.}
\beq
\label{part20}
\begin{array}{rcl}
{\cal L}_{aux.} &=& \displaystyle i(W_\Phi - \ov W_{\ov\Phi} ) f\ov f - \widetilde{m}^2 f
- \widetilde{M}^2 W_\Phi f - \ov{\widetilde{m}}^2 \ov f- \ov{\widetilde{M}}^2 \ov W_{\ov\Phi} \ov f 
\crbig
&& \displaystyle - {i\over2} W_{\Phi\Phi} [ \ov f\, \psi\psi - f \, \ov{\varphi\varphi} ]
+ {i\over2} \ov W_{\ov{\Phi\Phi}} [ f\,\ov{\psi\psi} - \ov f \, \varphi\varphi ] 
\,\,=\,\, - V + {\cal L}_{ferm.}.
\end{array}
\eeq
It generates the scalar potential
\beq
\label{part21}
V ={1\over i(W_\Phi - \ov W_{\ov\Phi})} \left| \widetilde{m}^2 + \widetilde{M}^2\,W_\Phi \right|^2.
\eeq
The term depending on $L$ in theory (\ref{part17}) does not contribute to the potential.
Fermion mass terms read
\beq
\label{part22}
\begin{array}{rcl}
{\cal L}_{ferm.} &=& \displaystyle 
-{1\over2} \widetilde{M}^2 \, W_{\Phi\Phi} \, \psi\psi
- {1\over2} \left[ \widetilde{m}^2 + \widetilde{M}^2\,W_\Phi \right] 
{W_{\Phi\Phi}\over W_\Phi - \ov W_{\ov\Phi}} \, \psi\psi + {\rm h.c.}
\crbig
&& \displaystyle - {1\over2} \left[ \widetilde{m}^2 + \widetilde{M}^2\,W_\Phi \right] 
{\ov W_{\ov{\Phi\Phi}}\over W_\Phi - \ov W_{\ov\Phi}} \, \varphi\varphi + {\rm h.c.}
\end{array}
\eeq
Three situations can occur. 

Firstly, if $\widetilde{M}^2=\widetilde{m}^2=0$, the theory has unbroken (linear) ${\cal N}=2$ supersymmetry and all fields are 
massless. This is also the case if $\widetilde{M}^2=0$, $\widetilde{m}^2\ne0$ and if the theory is canonical ({\it i.e.} free), 
$W_{\Phi\Phi}=0$, in which case the potential is an irrelevant constant $V \sim |\widetilde{m}|^4$.

Secondly, if the second supersymmetry is not deformed ($\widetilde{M}^2=0$), the theory is not free ($W_{\Phi\Phi}\ne0$) and
$\widetilde{m}^2\ne0$, ${\cal N}=2$ breaks to ${\cal N}=0$ with 
\beq
\label{part23}
\langle f\rangle = - {\ov{\widetilde{m}}^2 \over 2\Im \langle W_\Phi \rangle} .
\eeq
The theory has a vacuum state if $\langle W_{\Phi\Phi}\rangle=0$ has a solution, fermions remain then 
massless and the splitting of scalar masses is controlled by $\langle W_{\Phi\Phi\Phi}\rangle$.
This is also the case if $\widetilde{m}^2=0$ and $\widetilde{M}^2\ne0$ with 
\beq
\label{part24}
\langle f\rangle = - {\ov{\widetilde{M}}^2 \langle W_\Phi\rangle\over 2\Im \langle W_\Phi \rangle} .
\eeq

Thirdly, partial breaking to ${\cal N}=1$ occurs if $\widetilde{M}^2\ne0\ne \widetilde{m}^2$ and if the theory is not canonical 
($W_{\Phi\Phi}\ne0$). At the vacuum state,
\beq
\label{part25}
\langle W_\Phi \rangle = -{\widetilde{m}^2\over \widetilde{M}^2} \qquad\qquad \langle f \rangle =0 .
\eeq
Positivity of kinetic terms requires $\Im\langle W_\Phi\rangle < 0$.
The linear superfield $L$ remains of course massless, while the mass\footnote{Normalized with the metric
$-2\Im \langle W_\Phi\rangle$.} of $\Phi$ is controlled by 
$\langle W_{\Phi\Phi}\rangle$:
\beq
\label{part26}
{\cal M}^2_\Phi = \widetilde{M}^2\ov{\widetilde{M}}^2 \left| {\langle W_{\Phi\Phi}\rangle \over 2\Im \langle W_\Phi\rangle } \right|^2.
\eeq
In principle, $\Phi$ can acquire a very large mass and decouple from the massless $L$.

The analogy with partial supersymmetry breaking in a ${\cal N}=2$ Maxwell multiplet theory \cite{APT} is striking.
Describing this multiplet with ${\cal N}=1$ superfields $W_\alpha = - {1\over4}\ov{DD}D_\alpha V$ and $X$, with deformed supersymmetry variations
\beq
\label{Max0}
\delta^* W_\alpha =  -\sqrt2 M^2 \, \eta_\alpha + \sqrt 2 \, i \left[ \frac{1}{4}\eta_\alpha \ov{DD}\,\ov X 
+ i (\sigma^\mu\ov\eta)_\alpha \, \partial_\mu X \right] ,
\quad
\delta^*X = \sqrt 2 \, i \, \eta^\alpha W_\alpha ,
\eeq
the invariant lagrangian is written as
\beq
\label{Max2}
{\cal L}_{Max.} = {1\over2}\Fint \left[ {1\over2}{\cal F}_{XX}WW - {1\over4}{\cal F}_X
\ov{DD}\,\ov X + m^2 X -i M^2{\cal F}_X \right] + {\rm h.c.} + {\cal L}_{F.I.} ,
\eeq
where ${\cal F}(X)$ is the holomorphic prepotential and ${\cal L}_{F.I.} = \xi\int d^4\theta\, V$ is the 
Fayet-Iliopoulos (FI) term. Partial breaking arises if the theory is interacting, ${\cal F}_{XXX}\ne0$,
if $M^2\ne0\ne m^2$ and $\xi=0$. If we now compare with the lagrangian (\ref{part17}) and the deformed variation
\beq
\delta^*\ov D_\dalpha L = - \sqrt2 \,\widetilde{M}^2 \ov\eta_\dalpha + \sqrt2i \left[ {1\over4}\ov\eta_\dalpha \ov{DD}\ov \Phi
- i (\eta\sigma^\mu)_\dalpha\,\partial_\mu\Phi \right]\,,
\eeq
we observe that there is clearly a correspondence between $\Phi$ and $X$, ${\cal F}_X(X)$ and $W(\Phi)$ with a Lorentz chirality
inversion from $W_\alpha$ to $\ov D_\dalpha L$. However, there are significant differences, namely the absence of auxiliary fields in $L$ as well as the consequent inexistence of a corresponding ``electric'' FI term analogous to the $\xi D$ term for the Maxwell multiplet.

\subsection{Dual hypermultiplet formulation}

The duality transformation from the single--tensor to the hypermultiplet formulation is a Legendre transformation 
in ${\cal N}=1$ superspace. Instead of expression (\ref{part7}), let us use 
\beq
\label{part27}
{\cal L}_{kin.} = \Dint\Bigl[  {\cal H}(V, \Phi, \ov\Phi) - (S+\ov S)V \Bigr].
\eeq
The field equation for $S$ implies $V=L$ and the field equation for $V$ yields 
\beq
\label{part28}
{\cal H}_V = S+\ov S \quad \makebox{(Legendre transformation) }
\eeq
which allows one to express $V$ as a function of $S+\ov S$, $\Phi$ and $\ov\Phi$. The K\"ahler potential for the hypermultiplet 
with superfields $S$ and $\Phi$ is then
\beq
\label{part29}
{\cal K}(S+\ov S,\Phi,\ov\Phi) = \Bigl[{\cal H}(V,\Phi,\ov\Phi) - (S+\ov S)V \Bigr]_{V(S+\ov S,\Phi,\ov\Phi)}.
\eeq
In our case, the Legendre transformation is simply
\beq
\label{part30}
{\cal K}_V=0 \qquad\Longrightarrow\qquad S+\ov S = {\cal H}_V = -2iV(W_\Phi -\ov W_\Phi)
\eeq
with also
\beq
\label{part30b}
{\cal H}_S=0 \qquad\Longrightarrow\qquad {\cal K}_S = -V\,.
\eeq

The dual hypermultiplet theory reads
\beq
\label{part31}
\begin{array}{rcl}
{\cal L}_{dual} &=& \displaystyle 
i\Dint \left[ - {1\over4}
{(S+\ov S)^2 \over W_\Phi - \ov W_{\ov\Phi} }+ W\ov\Phi - \ov W\Phi \right]
+ \Fint \Bigl[ \widetilde{m}^2\Phi + \widetilde{M}^2 \, W \Bigr]
+ {\rm h.c.}
\crbig
&=& \displaystyle
\Fint \left[ -{i\over2} W_\Phi (\ov D{\cal K}_S)(\ov D{\cal K}_S) - {i\over4}W\ov{DD}\ov\Phi 
+ \widetilde{m}^2\Phi + \widetilde{M}^2 W \right] + {\rm h.c.}
\end{array}
\eeq
The $D$--term in the first expression is the K\"ahler potential of a hyper--K\"ahler space, $\det {\cal K}_{m\ov n} = 1/2$.
Since the superpotential depends on $\Phi$ only, the auxiliary component $f_S$ of $S$ does not contribute to
the potential. Its field equation
\beq
\label{part32}
(W_\Phi-\ov W_{\ov\Phi}) f_S - ( S+\ov S) W_{\Phi\Phi} f_\Phi = 0
\eeq
is actually the $\theta\theta$ component of the duality relation (\ref{part30}). The ground state in the partially 
broken phase is again characterized by relations (\ref{part25}) with, in addition, $\langle f_S\rangle=0$. On-shell, relations (\ref{part30}) and (\ref{part30b}) with $L$ replacing $V$, 
\beq
\label{part33}
{\cal H}_L = S+\ov S\,, 
\qquad\qquad\qquad
{\cal K}_S = -L\,,
\eeq
are consistent using the field equations for $L$ and $S$,
\beq
\label{part34}
\ov{DD}D_\alpha{\cal H}_L=0\,, \qquad\qquad \ov{DD}{\cal K}_S=0 \,,
\eeq
as integrability conditions.

That the ${\cal N}=1$ theory (\ref{part31}) has a second supersymmetry is not obvious. Since
the K\"ahler potential ${\cal K}$ generates a hyper--K\"ahler metric, the first term certainly has (on-shell) 
${\cal N}=2$ \cite{AGF}. Following~\cite{LR}, one easily verifies that ${\cal K}$ is invariant (up to a 
superspace derivative) under the variations
\beq
\label{part35}
\delta^*{\cal K}_S = {i\over\sqrt2}(\eta D\Phi + \ov{\eta D}\ov\Phi)\,,
\qquad\quad
\delta^*\Phi = -\sqrt2i\,\ov{\eta D} \, {\cal K}_S\,,
\qquad\quad
\delta^*\ov\Phi = -\sqrt2i\,\eta D \, {\cal K}_S\,,
\eeq
where ${\cal K}_S = {\partial\over\partial S}{\cal K} = -{i\over2} {S+\ov S \over W_\Phi- \ov W_{\ov\Phi}}$. These
variations are simply obtained by inserting the second duality relation (\ref{part33}) in the single--tensor off--shell 
variations (\ref{part3}). The field equation $\ov{DD}\,{\cal K}_S=0$ provides the linearity and chirality of
$\delta^*{\cal K}_S$ and $\delta^*\Phi$ respectively. The superpotential term $\widetilde{m}^2\Phi$ is also invariant.
The nonlinear deformation which allows for the presence of the superpotential $\widetilde{M}^2W$ is then
\beq
\label{part36}
\delta^*_{nl} \, \ov D_\dalpha{\cal K}_S = \sqrt2\, \widetilde{M}^2\, \ov\eta_\dalpha\,,
\eeq
in agreement with eqs.~(\ref{part15}) and (\ref{part33}).

\subsection{Several single-tensor multiplets}

The extension to a theory with several single--tensor multiplets is straigthforward. 
Consider the deformed ${\cal N}=2$ chiral superfields
\beq
{\cal Z}^a = \Phi^a + \sqrt2i \, \ov{\widetilde\theta D} L^a - {1\over4}\,\ov{\widetilde\theta\widetilde\theta}\, 
\Bigl[ 4 i (\widetilde{M}^a)^{2} + \ov{DD}\,\ov\Phi^a\Bigr]\,.
\eeq
The lagrangian
\beq
\begin{array}{rcl}
{\cal L} &=&  \Fint \bigint d^2\ov{\widetilde\theta}\, {\cal G}({\cal Z}^a) + {\rm h.c.}
\crbig
&=& \displaystyle
\Fint \Bigl[ {1\over2} {\cal G}_{ab} (\ov DL^a)(\ov DL^b) - {1\over4} {\cal G}_a \ov{DD} \ov \Phi^a -i (\widetilde{M}^a)^{2} \, {\cal G}_a + \widetilde{m}_a^2 \, \Phi^a
\Bigr] + {\rm h.c.}\,,
\end{array}
\eeq
where 
$$
{\cal G}_a = {\partial\over\partial\Phi^a} {\cal G}(\Phi^c)\,, \qquad\qquad
{\cal G}_{ab} = {\partial^2\over\partial\Phi^a\partial\Phi^b} {\cal G}(\Phi^c)\,,
$$
is invariant under the nonlinear second supersymmetry variations
\beq
\delta^* L ^a=  \sqrt2(\widetilde{M}^a)^{2} (\theta\eta+\ov{\theta\eta}) -{i\over\sqrt2} (\eta D\Phi+\ov{\eta D}\ov\Phi),
\quad\qquad
\delta^* \Phi^a = \sqrt2i\, \ov{\eta D}L^a .
\eeq
For $\widetilde{m}_a^2 \ne0\ne (\widetilde{M}^b)^{2}$, the condition for unbroken ${\cal N}=1$ is the cancellation of all auxiliary fields $f^a$: 
\beq
-i\langle {\cal G}_{ab} \rangle (\widetilde{M}^b)^2 + {\widetilde{m}}_a^2 = 0.
\eeq
In this vacuum, the kinetic metric $2\langle\Re {\cal G}_{ab}\rangle$ must be invertible and the mass matrix of the chiral multiplets
$\Phi^a$ is then
\beq
{\cal M}_{ab} = -{i\over2}\langle \Re {\cal G}_{ac} ^{-1}\rangle \langle{\cal G}_{bcd}\rangle (\widetilde{M}^d)^2,
\eeq
controlled by the third derivatives of ${\cal G}$.

\section{Nonlinear deformations}
\setcounter{equation}{0}

In the previous Section, we made use of particular nonlinear deformations of the ${\cal N}=2$ single--tensor and Maxwell 
multiplets to engineer theories with partial supersymmetry breaking. 
As illustrated by eq.~(\ref{part18}), a nonlinear deformation of the single--tensor multiplet can be introduced 
as a spurious constant component inserted in a ${\cal N}=2$ superfield. In this Section, we study general nonlinear
deformations of these multiplets, using their representation as chiral superfields in ${\cal N}=2$ superspace.

\subsection{Deformations of the Maxwell superfield}

A chiral--chiral (CC) ${\cal N}=2$ superfield describes the Maxwell multiplet:
\beq
\label{Max1}
{\cal W} (y,\theta,\widetilde\theta)
= X + \sqrt2i\,\widetilde\theta W - {1\over4}\widetilde\theta\widetilde\theta \,\ov{DD}\,\ov X ,
\qquad\qquad
\ov D_\dalpha{\cal W} = \ov{\widetilde D}_\dalpha{\cal W} =0,
\eeq
using chiral coordinates $\ov D_\dalpha \, y^\mu = \ov{\widetilde D}_\dalpha \, y^\mu = 0$, with also
\beq
\begin{array}{rcl}
W_\alpha  &=& -i\lambda_\alpha + \theta_\alpha D 
- {i\over2}(\sigma^\mu\ov\sigma^\nu\theta)_\alpha F_{\mu\nu}
- \theta\theta\, (\sigma^\mu\partial_\mu\ov\lambda)_\alpha ,
\crbig
X &=& x + \sqrt2\,\theta\kappa - \theta\theta\,F,
\crbig
{1\over4}\ov{DD}\,\ov X &=& \ov F + \sqrt2i\,\theta\sigma^\mu\partial_\mu\ov\kappa + \theta\theta\,\Box\ov x.
\end{array}
\eeq
The $SU(2)_R$ symmetry of the ${\cal N}=2$ algebra acts linearly on the components of the Maxwell 
superfield ${\cal W}$. Defining fermion doublets
\beq
\theta^1 = \theta, \qquad \theta^2=\widetilde\theta, \qquad \lambda_1=\kappa, \qquad \lambda_2=\lambda, \quad
\eeq
leads to
\beq
{\cal W}(y,\theta,\widetilde\theta) = x + \sqrt2\, \theta^i\lambda_i - \theta^i\theta^j Y_{ij} + \ldots
\eeq
omitting terms which depend on derivatives of the fields. 
Since $\theta^i\theta^j = \theta^j\theta^i$, 
\beq
Y_{ij}=Y_{ji} = [ \vec Y\cdot\vec\sigma \, \sigma_2]_{ij}
\eeq 
and the vector $\vec Y$ is in general a complex $SU(2)_R$ triplet.
But in ${\cal W}$, the auxiliary fields correspond to
\beq
\label{Yis1}
Y_{11} = F, \qquad Y_{22} = \ov F, \qquad Y_{12} = -{i\over\sqrt2}\, D, \qquad  
\vec Y = \left( \Im F, \Re F, {D\over\sqrt2} \right)
\eeq
and the $SU(2)_R$--invariant ``reality" condition
\beq
\label{Yis2}
Y^{ij} \equiv Y_{ij}^* = \epsilon^{ik}\epsilon^{jl}Y_{kl}
\eeq
is verified: a complex value of $\vec Y$ violating this condition cannot be seen as a background value of 
${\cal N}=1$ superfields $X$ or $W_\alpha$.

Since gauginos are in the $\theta^i$ components, nonlinear deformations of their variations, as expected 
for goldstino fermions, should be introduced with
\beq
\label{Yis3}
\begin{array}{rcl}
{\cal W}_{nl} &=& A^2 \theta\theta + B^2 \widetilde\theta\widetilde\theta+ 2 \Gamma \theta\widetilde\theta , \qquad\qquad\makebox{($A$, $B$, $\Gamma$ complex)}
\crbig
&=& (Y_2+iY_1) \, \theta\theta + (Y_2-iY_1) \, \widetilde\theta\widetilde\theta - 2iY_3 \, \theta\widetilde\theta
\end{array}
\eeq
added to ${\cal W}$. Then, $\vec Y = \left( -{i\over2}[A^2-B^2] , {1\over2}[A^2+B^2] , i\Gamma \right)$ and
\beq
\delta\kappa_\alpha = \sqrt2( A^2 \epsilon_\alpha + \Gamma \eta_\alpha) + \ldots
\qquad\qquad
\delta\lambda_\alpha =  \sqrt2( B^2 \eta_\alpha + \Gamma \epsilon_\alpha) + \ldots
\eeq
If $\Gamma=\pm AB$, ${\cal W}_{nl} = (A\theta \pm B\widetilde\theta)^2$,
$\delta(B\kappa_\alpha \mp A\lambda_\alpha)=0$ and the deformation partially breaks ${\cal N}=2$ to 
${\cal N}=1$. We earlier used the particular case $A=\Gamma=0$. The condition for partial breaking is in any case
incompatible with the reality condition (\ref{Yis2}): the auxiliary fields $F$ and $D$ are not able to induce partial 
breaking with their background values; in other words, the deformation parameters cannot be absorbed in the background values of the auxiliary fields, in contrast with the case of the spontaneous breaking of $\mathcal{N}=1$.
An $SU(2)$ rotation can be used to cancel $Y_3=i\Gamma$. With this choice, partial breaking occurs either if
$A=0$, and the goldstino is $\lambda_\alpha$, or if $B=0$ and the goldstino is $\kappa_\alpha$.

The condition for partial breaking has an elegant $SU(2)$-invariant formulation in terms of the complex vector 
$\vec Y$: we need $\vec Y\ne 0$ (to break) and $\vec Y\cdot\vec Y=0$ (to {\it partially} break). Hence,
$\vec Y$ must be complex, electric {\it and} magnetic.

\subsection{Deformations of the single--tensor superfield}

While a chiral--chiral (CC) superfield is relevant to study deformations of the Maxwell multiplet, 
the single--tensor multiplet is conveniently described using a chiral--antichiral (CA) ${\cal N}=2$ superfield 
${\cal Z}$,
\beq
\label{st1}
\ov D_\dalpha\,{\cal Z} = \widetilde D_\alpha\,{\cal Z}  = 0 \,,
\eeq
with the expansion
\beq
\label{st2}
{\cal Z} = \Phi + \sqrt2i \, \ov{\widetilde\theta D}L - {1\over4}\,\ov{\widetilde\theta\widetilde\theta}\, \ov{DD}\,\ov\Phi
\eeq
in the appropriate coordinates $(\widetilde y, \theta, \ov{\widetilde\theta})$, $\ov D_\dalpha\,\widetilde y^\mu 
= \widetilde D_\alpha\, \widetilde y^\mu = 0$. A particular deformation with partial supersymmetry breaking 
has been earlier described [eq.~(\ref{part19})] and we wish to generalize it.
Since fermion fields are in the components\footnote{The field components of $\Phi$ are $z$, $\psi$ and $f$
and $\ov D_\dalpha L$ is expanded in eq.~(\ref{part6b}).}
\beq
\label{st3}
\sqrt2 \, \theta\psi - \sqrt2 \, \ov{\widetilde \theta \varphi  }
\eeq
of ${\cal Z}$, the deformation parameters will add 
\beq
\label{st4}
{\cal Z}_{nl} = \widetilde{A}^2 \, \theta\theta + \ov{\widetilde{B}}^2 \,\ov{\widetilde\theta\widetilde\theta}
\eeq
to ${\cal Z}$. In contrast with the Maxwell case, the mixed contribution $\theta_\alpha \ov{\widetilde\theta}_\dalpha$
is a space--time vector and the deformations are encoded in two complex numbers $\widetilde{A}^2$ and 
$\widetilde{B}^2$ only. The nonlinear variations of the spinors are 
\beq
\label{st5}
\delta\psi_\alpha = \sqrt2 \, ( \widetilde{A}^2 - f) \, \epsilon_\alpha  + \ldots
\qquad\qquad
\delta\varphi_\alpha =  - \sqrt2 \, ( \widetilde{B}^2 + f) \, \eta_\alpha  + \ldots
\eeq
and generic values of $\widetilde{A}^2$ and $\widetilde{B}^2$ break both supersymmetries. 
Partial breaking occurs if either $\widetilde{B}^2=0$ and the goldstino is $\psi$ in $\Phi$, or if $\widetilde{A}^2=0$ with $\varphi$ 
in $L$ as the goldstino.
An expectation value $\langle f \rangle$ of the auxiliary $f$ in $\Phi$ corresponds to $\widetilde{A}^2= -\widetilde{B}^2$ and cannot generate partial breaking on its own. 

In the linear ${\cal N}=2$ theory, all fields are massless since the single--tensor multiplet includes a 
tensor with gauge symmetry. A generic lagrangian generated by the CA superfield ${\cal Z}$ is
\beq
\label{st6}
{\cal L} = \Fint \left[ \bigint d^2\ov{\widetilde\theta}\, {\cal G}({\cal Z}) + \widetilde{m}^2\,\Phi \right]
+ {\rm h.c.} = {\cal L}_{lin.} + {\cal L}_{nl}\,.
\eeq
where ${\cal L}_{nl}$ includes all terms generated by the deformations with parameters $\widetilde{A}^2$ and $\widetilde{B}^2$.
In the function ${\cal G}({\cal Z})$, a term linear in ${\cal Z}$ is irrelevant (it contributes with a derivative) and
the component expansion of the lagrangian depends on the second and higher derivatives of ${\cal G}$.
The only auxiliary field is $f$ in $\Phi$ and ${\cal L}_{lin.}$ includes the terms
\beq
\label{st7}
[ {\cal G}^{\prime\prime}(z) + \ov{\cal G}^{\prime\prime}(\ov z) ] \, \ov f f
+ \left[ {1\over2} {\cal G}^{\prime\prime\prime}(z) [ \ov f \, \psi\psi + f \, \ov{\varphi\varphi} ] - \widetilde{m}^2\,f\right]
+ {\rm h.c.}
\eeq
The parameter $\widetilde{m}^2$ induces $\langle f\rangle = {\widetilde{m}^2 / 2\langle\Re {\cal G}^{\prime\prime}\rangle}$ which
breaks both supersymmetries if the theory is not canonical, $ {\cal G}^{\prime\prime\prime}\ne0$.
The nonlinear deformation produces the following terms:
\beq
\label{st8}
\begin{array}{rcl}
{\cal L}_{nl} &=& - {\cal G}^{\prime\prime}(z) \Bigl[ \ov{\widetilde{B}}^2\, f + \widetilde{A}^2 \, \ov f + \widetilde{A}^2\ov{\widetilde{B}}^2 \Bigr] + {\rm h.c.}
\crbig
&& \displaystyle - {1\over2} {\cal G}^{\prime\prime\prime}(z) \Bigl[ \ov{\widetilde{B}}^2\, \psi\psi +\widetilde{A}^2 \, \ov{\varphi\varphi} 
\Bigr] + {\rm h.c.}
\end{array}
\eeq
Hence, 
$$
2\,[\Re{\cal G}^{\prime\prime}(z)]\,\ov f = {\cal G}^{\prime\prime}(z) \ov{\widetilde{B}}^2  + \ov{\cal G}^{\prime\prime}(\ov z) \ov{\widetilde{A}}^2 
+ \widetilde{m}^2 - {1\over2} {\cal G}^{\prime\prime\prime}(z) \ov{\varphi\varphi} 
-{1\over2} \ov{\cal G}^{\prime\prime\prime}(\ov z) \ov{\psi\psi}  
$$
and the scalar potential and the fermion bilinear terms read respectively
\beq
\label{st9}
\begin{array}{rcl}
V (z,\ov z) &=& \displaystyle {1\over2\Re{\cal G}^{\prime\prime}} \Bigl| \widetilde{B}^2 \,\ov{\cal G}^{\prime\prime}
+ \widetilde{A}^2 \, {\cal G}^{\prime\prime} + {\ov{\widetilde{m}}}^2 \Bigr|^2 + 2\Re [ \widetilde{A}^2\ov{\widetilde{B}}^2 \, {\cal G}^{\prime\prime} ], 
\crbig
{\cal L}_{ferm.} &=& \displaystyle 
{1\over2} \psi\psi \Bigl[ {{\cal G}^{\prime\prime\prime}\over2\Re{\cal G}^{\prime\prime}} 
(\ov{\widetilde{B}}^2 \,{\cal G}^{\prime\prime} + \ov{\widetilde{A}}^2 \, \ov {\cal G}^{\prime\prime} + \widetilde{m}^2)
- \ov{\widetilde{B}}^2 \, {\cal G}^{\prime\prime\prime}
\Bigr] + {\rm h.c.}
\crbig
&& + \displaystyle {1\over2} \varphi\varphi \Bigl[ {\ov{\cal G}^{\prime\prime\prime}\over2\Re{\cal G}^{\prime\prime}}  
(\ov{\widetilde{B}}^2 \,{\cal G}^{\prime\prime} + \ov{\widetilde{A}}^2 \, \ov {\cal G}^{\prime\prime} +\widetilde{m}^2)
- \ov{\widetilde{A}}^2 \, {\cal G}^{\prime\prime\prime}
\Bigr] + {\rm h.c.} 
\end{array}
\eeq
The kinetic metric of the multiplet is $2 \Re{\cal G}^{\prime\prime}(z)$.
Notice that these formulas do not depend on the real scalar $C$ in $L$, which always leads to a flat direction. 

If $\widetilde{A}\widetilde{B}=0$ with ${\cal L}_{nl}\ne0$ and the ground state equation 
$\langle \ov{\widetilde{B}}^2 \,{\cal G}^{\prime\prime} + \ov{\widetilde{A}}^2 \, \ov {\cal G}^{\prime\prime} + \widetilde{m}^2 \rangle=0$
has a solution, one supersymmetry remains unbroken: $\langle f\rangle=0$. This requires $\widetilde{m}^2\ne0$, since
positivity of the kinetic metric forbids $\langle{\cal G}^{\prime\prime}\rangle=0$.
If $\widetilde{B}\ne0$, the mass terms are
$$
2\langle\Re{\cal G}^{\prime\prime}\rangle \Bigl[ \mathcal{M}_\Phi \ov{\mathcal{M}}_\Phi \, z \ov z - {1\over2}  \mathcal{M}_\Phi \psi\psi 
- {1\over2} \ov{\mathcal{M}}_\Phi\, \ov{\psi\psi} \Bigr],
\qquad\qquad
\mathcal{M}_\Phi = { \ov{\widetilde{B}}^2 \langle{\cal G}^{\prime\prime\prime}\rangle \over 2\langle\Re{\cal G}^{\prime\prime}\rangle} .
$$
This is the case already obtained in eqs.~(\ref{part25}) and (\ref{part26}): the chiral ${\cal N}=1$ superfield
$\Phi$ has mass $ \mathcal{M}_\Phi$, and $L$ is massless. 
If $\widetilde{A}\ne0$, the mass terms are
$$
2\langle\Re{\cal G}^{\prime\prime}\rangle \Bigl[  \mathcal{M}_\Phi \ov{\mathcal{M}}_\Phi \, z \ov z - {1\over2}  \mathcal{M}_\Phi \varphi\varphi 
- {1\over2} \ov{\mathcal{M}}_\Phi \, \ov{\varphi\varphi} \Bigr],
\qquad\qquad
 \mathcal{M}_\Phi = { \ov{\widetilde{A}}^2\langle{\cal G}^{\prime\prime\prime}\rangle \over 2\langle\Re{\cal G}^{\prime\prime}\rangle} .
$$
The roles of $\psi$ and $\varphi$ are exchanged, the ${\cal N}=1$ multiplet with mass $ \mathcal{M}_\Phi$ has fields $z$ and 
$\varphi$, while $\psi$ is the ${\cal N}=1$ partner of $H_{\mu\nu\rho}$ and $C$ in the massless linear 
superfield. 

If $\widetilde{A}\widetilde{B}\ne0$, the non--zero second term in the scalar potential (which can have both signs)
breaks both supersymmetries, assuming that $V$ has a ground state $\langle z\rangle$.

\section{Constrained multiplets}
\setcounter{equation}{0}

When supersymmetry is partially broken in the Maxwell or single--tensor (hypermultiplet) theory, a chiral multiplet ($X$ or $\Phi$) acquires an arbitrary mass. In the infinite--mass limit, the field equation of this superfield is a constraint which allows for the elimination of the massive chiral superfield. One is then left with a nonlinear realization of ${\cal N}=2$ supersymmetry in terms of the $4_B+4_F$ fields of the ${\cal N}=1$ Maxwell or linear superfield. 

\subsection{The infinite--mass limit}

We begin with partial breaking in the Maxwell theory. Since the two options $A^2=0$ and $B^2=0$ are equivalent, we only consider the first case and use the deformed chiral--chiral deformed superfield
\beq \label{firstw}
{\cal W} = X + \sqrt2i\,\widetilde\theta W + \widetilde\theta\widetilde\theta \, \left[ B^2
- {1\over4}\ov{DD}\,\ov X \right] \,,
\eeq
in terms of which the lagrangian is
\beq
\label{Mcon1}
\begin{array}{rcl}
{\cal L} &=& \displaystyle {1\over2} \Fint\left[ \bigint d^2\widetilde\theta\, {\cal F}({\cal W}) + m^2 X \right] + {\rm h.c.}
+ {\cal L}_{F.I.}
\crbig
&=& \displaystyle {1\over4}\Fint \left[ {\cal F}_{XX}WW - {1\over2}{\cal F}_X
\ov{DD}\,\ov X + 2m^2 X + 2 B^2{\cal F}_X \right] + {\rm h.c.} + {\cal L}_{F.I.}
\end{array}
\eeq
Since the auxiliary fields $f$ and $D$ vanish in the ground state, the mass terms of the fermion $\chi$ in $X$ are
$$
- {B^2\over4}\, \langle{\cal F}_{XXX}\rangle\, \chi\chi 
- {\ov B^2\over4}\, \langle\ov{\cal F}_{\ov{XXX}}\rangle \, \ov{\chi\chi }
$$
and, since the kinetic metric is $\Re \langle{\cal F}_{XX} \rangle$, the mass of $X$ is 
\beq
\label{Mcon2}
{\cal M}_X = {B^2 \, \langle{\cal F}_{XXX}\rangle \over 2\, \Re \langle{\cal F}_{XX} \rangle} .
\eeq
The infinite--mass limit is $\langle{\cal F}_{XXX}\rangle \rightarrow \infty$ with 
fixed $\Re \langle{\cal F}_{XX} \rangle$ (as the latter corresponds to the metric of the scalar manifold), thus disproving the claim made in \cite{RT}. Expanding the field equation of $X$ and retaining only the term
in $\langle{\cal F}_{XXX}\rangle$ leads to the constraint
\beq
\label{Mcon3}
WW - {1\over 2} X\ov{DD}\ov X + 2\, B^2 \,X = 0\,,
\eeq
which was first given in \cite{BG}. Multiplying (\ref{Mcon3}) by $W_\alpha$ or $X$ leads also to $XW_\alpha=X^2=0$ and the constraint (\ref{Mcon3}) is then equivalent to \cite{ADM}
\beq
\label{Mcon4}
{\cal W}^2 = 0\,.
\eeq 

We now turn to the partial breaking in a single--tensor theory. Again, the two options $\widetilde{A}^2=0$ and $\widetilde{B}^2=0$ are equivalent, so we only consider the first case and use the deformed 
chiral--antichiral superfield 
\beq \label{firstz}
{\cal Z} = \Phi + \sqrt2i \, \ov{\widetilde\theta D} L + \ov{\widetilde\theta\widetilde\theta}\, \left[ \ov{\widetilde{B}}^2
- {1\over4}\,\ov{DD}\,\ov\Phi \right] \,,
\eeq
which induces the nonlinear deformation
\beq
\delta^*_{nl} \, \ov D_\dalpha L \,\,=\,\, - i \sqrt2\, \ov{\widetilde{B}}^2 \, \ov\eta_\dalpha \,.
\eeq
The theory (\ref{st6}) and the field equation for $\Phi$ respectively read 
\beq
\label{st10} 
\begin{array}{rcl}
{\cal L} &=& \displaystyle \Fint\left[ {\cal G}_\Phi (\Phi) \left( -{1\over4}\ov{DD}\ov\Phi + \ov{\widetilde{B}}^2 \right)
+ {1\over2}{\cal G}_{\Phi\Phi}(\Phi)(\ov DL)(\ov DL) + \widetilde{m}^2\Phi\right] + {\rm h.c.}\,,
\crbig
0 &=& {\cal G}_{\Phi\Phi}(\Phi) \left( -{1\over4}\ov{DD}\ov\Phi + \ov{\widetilde{B}}^2 \right)
+ {1\over2}{\cal G}_{\Phi\Phi\Phi}(\Phi)(\ov DL)(\ov DL) + \widetilde{m}^2.
\end{array}
\eeq
The lowest component is the field equation for the auxiliary field $f$, 
$$
{\cal G}_{zz}(z)\,(\ov f-\ov{\widetilde{B}}^2) =  \widetilde{m}^2
$$
omitting fermions, and $\langle f \rangle=0$ defines the ground state
${\cal G}_{zz}(\langle z \rangle) = - \widetilde{m}^2/\ov{\widetilde{B}}^2$ and the kinetic metric normalization
$2\Re {\cal G}_{zz}(\langle z \rangle)$. 

As explained earlier, the mass of $\Phi$ is
controlled by ${\cal G}_{zzz}(\langle z \rangle)$ and this free parameter can be sent 
to infinity keeping ${\cal G}_{zz}(\langle z \rangle)$ finite as in the Maxwell case. 
In this limit,
$$
{\cal G}_{zz}(\Phi) \sim {\cal G}_{zzz}(\langle z \rangle) [\Phi - \langle z\rangle],
\qquad\qquad  
{\cal G}_{zzz}(\Phi) \sim  {\cal G}_{zzz}(\langle z \rangle)
$$
and the field equation becomes\footnote{One can redefine $\Phi-\langle z\rangle \longrightarrow \Phi$.} 
\beq
\label{st11}
{1\over2} \Phi  \ov{DD}\ov\Phi -(\ov DL)(\ov DL) = 2\ov{\widetilde{B}}^2\, \Phi\,,
\eeq
which does not depend on the function ${\cal G}$ and which was first given in \cite{Bagger:1997pi}. This equation allows to eliminate $\Phi$. The solution expresses $\Phi$ as
a function of $(\ov DL)(\ov DL)$, with
\beq
\label{st12} 
\Phi = - {2(\ov DL)(\ov DL) \over 4\ov{\widetilde{B}}^2 - \ov{DD}\ov\Phi}  \qquad\Longrightarrow\qquad
\Phi\, \ov D_\dalpha L = \Phi^n=0 \quad (n\ge2).
\eeq
The second supersymmetry variation of the constraint (\ref{st11}) is 
\beq
\delta^* \, \Big[ {1\over2} \Phi  \ov{DD}\ov\Phi -(\ov DL)(\ov DL) - 2\ov{\widetilde{B}}^2\, \Phi \Big] = -2 \sqrt2 \, \partial_\mu (\eta \sigma^\mu \ov{D}L \Phi)\,.
\eeq
The invariance of the constraint then follows from the results (\ref{st12}). Moreover, since
\beq
\label{st13} 
{\cal Z}^2 = \Phi^2 + 2\sqrt2i \, \Phi \, \ov{\widetilde\theta D} L 
- \ov{\widetilde\theta\widetilde\theta} \, \Bigl[ {1\over2} \Phi  \ov{DD}\ov\Phi  -(\ov DL)(\ov DL) - 2\ov{\widetilde{B}}^2 \, \Phi 
\Bigr] ,
\eeq
eq.~(\ref{st11}) is equivalent to the ${\cal N}=2$ condition
\beq
\label{st14}
{\cal Z}^2 = 0 \, .
\eeq

\subsection{Solutions of the constraints}

The solution of (\ref{Mcon3}), and thus of (\ref{Mcon4}), was first given in \cite{BG}. In our conventions, it is
\beq \label{opk}
X = - \frac{W^2}{2B^2} \Bigg[ 1- \ov{D}^2 \Bigg( \frac{\ov{W}^2}{4B^4 +a+4B^4 \sqrt {1+ \frac{a}{2B^4} + \frac{b^2}{16B^8}} } \Bigg) \Bigg] \,,
\eeq
where
\beq
a = \frac{1}{2} (D^2W^2+\ov{D}^2\ov{W}^2) \, , \, b = \frac{1}{2} (D^2W^2-\ov{D}^2\ov{W}^2) \,.
\eeq
The bosonic part of lagrangian (\ref{Mcon1}) then takes the form
\beq
\label{Sollag}
\begin{array}{rcl}
{\cal L}|_{bos} &=&  8m^2 B^2 \Big(  1- \sqrt{1-\frac{1}{B^4} (-F_{\mu \nu} F^{\mu \nu}  +2D^2) - \frac{1}{4B^8} (F_{\mu \nu} \tilde{F}^{\mu \nu})^2}\, \Big) \,.
\end{array}
\eeq
The equation of motion for $D$ is then
\beq
D=0\,,
\eeq
and, substituting back into (\ref{Sollag}), one arrives at \cite{BG}, \cite{ADM}
\beq
\label{Sollag2}
\begin{array}{rcl}
{\cal L}|_{bos} &=&  8m^2 B^2 \Big(  1- \sqrt{1+\frac{1}{B^4} F_{\mu \nu} F^{\mu \nu} - \frac{1}{4B^8} (F_{\mu \nu} \tilde{F}^{\mu \nu})^2}\, \Big)
\crbig
&=& 8m^2 B^2 \Big(  1- \sqrt{-\det \big(\eta_{\mu \nu} - \frac{\sqrt2}{B^2}F_{\mu \nu} \big)}\, \Big)\,.
\end{array}
\eeq

It is also possible to add the FI term 
\beq
\xi \Dint V = \frac{1}{2} \xi D
\eeq
to the lagrangian (\ref{Sollag}). Solving the equation of motion for $D$ then gives
\beq
\begin{array}{rcl}
- \frac{2}{B^4} D^2 &=& - \frac{\xi^2}{\xi^2 + 2 \cdot 16^2 m^4}  \Big(1+\frac{1}{B^4} F_{\mu \nu} F^{\mu \nu} - \frac{1}{4B^8} (F_{\mu \nu} \tilde{F}^{\mu \nu})^2 \Big)\,,
\end{array}
\eeq
and substituting back to (\ref{Sollag}), we find that the latter takes the form
\beq
\begin{array}{rcl}
\mathcal{L}|_{bos} &=& 8m^2 B^2 \Bigg(  1- \Big(1+ \frac{\xi^2}{4 \cdot 8 \cdot 16 m^4} \Big)\,\sqrt{1-\frac{1}{B^4} (-F_{\mu \nu} F^{\mu \nu}  +2D^2) - \frac{1}{4B^8} (F_{\mu \nu} \tilde{F}^{\mu \nu})^2}\, \Bigg)
\crbig
&=& 8m^2 B^2 \Big(  1- \sqrt{1+\frac{\xi^2}{8^3 m^4}} \,\sqrt{-\det \big(\eta_{\mu \nu} - \frac{\sqrt2}{B^2}F_{\mu \nu} \big)}\, \Big)\,,
\end{array}
\eeq
which means that the addition of the FI term only changes the prefactor of the Born--Infeld lagrangian included in $\mathcal{L}$.

Following \cite{BG}, \cite{Bagger:1997pi} and \cite{AADT}, we now give the solution $\Phi=\Phi(\ov{D}L)$ of the constraint (\ref{st12}) or equivalently of (\ref{st14}). In our conventions, it is 
 \beq
\label{asc}
\begin{array}{rcl}
{\Phi} =  - \frac{1}{2 \widetilde{B}^2} \Big[ (\ov{D}L)^2 - \ov{D}^2 \Bigg( \frac{(DL)^2 (\ov{D}L)^2}{4 \widetilde{B}^4 + \widetilde{a} + 4\widetilde{B}^4 \sqrt{1 + \frac{\widetilde{a}}{2\widetilde{B}^4}  + \frac{\widetilde{b}^2}{16\widetilde{B}^8}  }}\Bigg) \Big] \,,
\end{array}
\eeq
where we have assumed that $\widetilde{B}$ is real for simplicity and
\beq
\widetilde{a} = \frac{1}{2}\Big(\ov{D}^2 [(DL)^2] + D^2 [(\ov{D}L)^2]\Big) = \ov{\widetilde{a}} \, , \, \widetilde{b}=\frac{1}{2} \Big(\ov{D}^2 [(DL)^2] - D^2 [(\ov{D}L)^2]\Big) = -\ov{\widetilde{b}} \,.
\eeq
 Due to the constraint (\ref{st14}), only if $\mathcal{G}$ has linear dependence on $\mathcal{Z}$ will it contribute to (\ref{st6}). However,
\beq
\Fint  d^2 \ov{\tilde{\theta}} \, \mathcal{Z} + \textrm{h.c.} \sim \Fint  \Big(  \widetilde{B}^2
- {1\over4}\,\ov{DD}\,\ov\Phi \Big) + \textrm{h.c.}  = \textrm{derivative} \,.
\eeq
Consequently, (\ref{st6}) takes the form
\beq
\label{pfg}
\begin{array}{rcl}
{\cal L} &=&  \displaystyle \widetilde{m}^2 \Fint \Phi + {\rm h.c.} 
\crbig
&=& - \frac{\widetilde{m}^2}{2\widetilde{B}^2} \Fint \, (\ov{D}L)^2 \Bigg[1- \ov{D}^2 \Bigg(  \frac{(DL)^2}{4 \widetilde{B}^4 + \widetilde{a} + 4\widetilde{B}^4 \sqrt{1 + \frac{\widetilde{a}}{2\widetilde{B}^4}  + \frac{\widetilde{b}^2}{16\widetilde{B}^8}  }} \Bigg) \Bigg]+ {\rm h.c.} 
\end{array}
\eeq
Moreover, using (\ref{part6b}), we find
\beq
\begin{array}{ccc}
(\ov{D}L)^2|_{bos}&=& \theta^2 \, (\upsilon_\mu \upsilon^\mu + 2i \upsilon_\mu \partial^\mu C - \partial_\mu C \partial^\mu C  ) \,,
\crbig
\widetilde{a}|_{ bos} &=& 4\, \big(\upsilon^2 - (\partial C)^2\big) \quad , \quad \widetilde{b}|_{bos} = -8i \, \upsilon \cdot \partial C \,.
\end{array}
\eeq
Then
\beq \label{dbian}
\begin{array}{rcl}
\mathcal{L}|_{bos} &=& \widetilde{m}^2 \widetilde{B}^2 \Big(1- \sqrt{1+ \frac{2}{\widetilde{B}^4} \big(\upsilon^2 - (\partial C)^2\big)  - \frac{4}{\widetilde{B}^8}(\upsilon \cdot \partial C)^2} \,\Big) 
\crbig
&=& \widetilde{m}^2 \widetilde{B}^2 \Bigg(1- \sqrt{1- \frac{2}{\widetilde{B}^4} \Big(\frac{1}{6}H_{\mu \nu \rho} H^{\mu \nu \rho} + \partial_\mu C \partial^\mu C \Big)  - \frac{1}{9\widetilde{B}^8}(\epsilon_{\mu \nu \rho \sigma} H^{ \nu \rho \sigma} \partial^{\mu} C)^2} \,\Bigg) \,.
\end{array}
\eeq

\section{The ``long" super-Maxwell superfield}
\setcounter{equation}{0}

In Section 6 we will construct supersymmetric interactions of deformed or constrained single--tensor and Maxwell supermultiplets. We will find it useful to describe the Maxwell multiplet in terms of a chiral--antichiral
superfield, with $16_B+16_F$ components, as an alternative to the $8_B+8_F$ chiral--chiral
superfield (\ref{Max1}). In the present and technical Section, we thus proceed to construct this ``long" ${\cal N}=2$
superfield for the super--Maxwell theory.

To begin with, both types of superfields exist for the single--tensor multiplet. In particular, the latter can be described either by the ``short" ($8_B+8_F$) chiral--antichiral 
(CA) superfield (\ref{st2}),
\beq
\label{longM0}
{\cal Z} =  \Phi + \sqrt2 i\, \ov{\widetilde\theta D} \, L
- \ov{\widetilde\theta\widetilde\theta}\, {1\over4} \, \ov{DD} \, \ov\Phi ,
\eeq
(and its AC conjugate), 
or by a ``long" chiral--chiral (CC) superfield  \cite{AADT}
\beq
\label{longM1}
\widehat{\cal Z} = Y + \sqrt 2 \, \widetilde\theta \chi - \widetilde\theta\widetilde\theta\, \Bigl[ {i\over2}\Phi 
+ {1\over4}\ov{DD} \, \ov Y \Bigr] ,
\eeq
where $Y$, $\Phi$ and $\chi_\alpha$ are chiral ${\cal N}=1$ superfields with $16_B+16_F$ field components.
They are related by\,\footnote{Identities in Apprendix \ref{AppA} may help.}
\beq
\label{longM2}
{\cal Z} =  -{i\over2}\, \widetilde D\widetilde D \, \widehat{\cal Z} + {i\over2}\, \ov{DD} \, \ov{\widehat{\cal Z}}
\eeq
and the real linear superfield $L$ is 
\beq
\label{longM3}
L = D\chi -\ov{D\chi}.
\eeq
Chirality of $\chi_\alpha$ implies linearity of $L$. 

There is a gauge invariance acting on the long CC superfield. According to eqs. (\ref{longM0}) and
(\ref{longM3}), ${\cal Z}=0$ if $\Phi=0$ and $D\chi =\ov{D\chi}$. The second condition is a Bianchi identity verified by
\beq
\label{longM4}
\chi_\alpha = -{i\over4}\ov{DD}D_\alpha \Pi, \qquad\qquad \ov\chi_\dalpha = -{i\over4}DD\ov D_\dalpha \Pi, 
\qquad\qquad 
\makebox{($\Pi$ real)}.
\eeq
Hence, ${\cal Z}$ is invariant under 
\beq
\label{longM5}
\widehat{\cal Z} \qquad\longrightarrow\qquad \widehat{\cal Z} + {\cal W}
\eeq
where ${\cal W}$ is a Maxwell (chiral--chiral) superfield (\ref{Max1}). This gauge invariance eliminates $8_B+8_F$
components in $\widehat{\cal Z}$. We now proceed to construct a ``long" chiral--antichiral ${\cal N}=2$
superfield for the super--Maxwell theory.

\subsection[The chiral--antichiral ${\cal N}=2$ superfield]{The chiral--antichiral \boldmath{${\cal N}=2$} superfield} \label{sec51}

A generic chiral--antichiral superfield, $\ov D_\dalpha\widehat{\cal W} = \widetilde D_\alpha
\widehat{\cal W}=0$, 
has the expansion
\beq
\label{CA1}
\widehat{\cal W} = U + \sqrt2\, \ov{\widetilde\theta \, \Omega} - \ov{\widetilde\theta\widetilde\theta}\,\Bigl[
{i\over2}X +{1\over4}\ov{DD}\,\ov U \Bigr] ,
\eeq
where the ${\cal N}=1$ superfields $U, X$ and $\ov\Omega_\dalpha$ which include $16_B+16_F$ fields, 
are chiral: they vanish under $\ov D_\dalpha$.  In components, $\ov\Omega_\dalpha$ 
includes a complex vector $\mathbb{V}_\mu$ ($8_B$) and two Majorana fermions:
\beq
\label{CA2}
\ov\Omega_\dalpha = \ov\omega_\dalpha + (\theta\sigma^\mu)_\dalpha\,\mathbb{V}_\mu
- \theta\theta\,\ov\lambda_\dalpha.
\eeq
Such a chiral right--handed (the index $\dalpha$) spinor superfield can always be written as
\beq
\label{CA3}
\ov\Omega_\dalpha = \ov D_\dalpha\,\mathbb{L}, \qquad\qquad
\Omega_\alpha = - D_\alpha\,\ov\mathbb{L}, \qquad\qquad
\eeq
where $\mathbb{L}$ is complex linear, $\ov{DD}\,\mathbb{L}=0$. In components, a complex
linear superfield can be written
\beq
\label{CA4}
\mathbb{L} (x,\theta,\ov\theta) = \Phi (x,\theta,\ov\theta) - \ov{\theta\omega} 
- \theta\sigma^\mu\ov\theta\,\mathbb{V}_\mu
+ \theta\theta\ov{\theta\lambda} - {i\over2}\,\ov{\theta\theta}\,\theta\sigma^\mu\partial_\mu\ov\omega
+ {i\over2}\theta\theta\ov{\theta\theta} \, \partial^\mu\mathbb{V}_\mu
\eeq
with $\Phi$ chiral, $\ov D_\dalpha\Phi=0$, an expansion which leads directly to $\ov D_\dalpha\,\mathbb{L}
= \ov\Omega_\dalpha$ in eq.~(\ref{CA2}). In other words, 
\beq
\label{CA5}
\widehat{\cal W} = U + \sqrt2\, \ov{\widetilde\theta \, D} \, \mathbb{L} - \ov{\widetilde\theta\widetilde\theta}\,\Bigl[
{i\over2}X +{1\over4}\ov{DD}\,\ov U \Bigr] 
\eeq
in general. 

Upon defining the chiral--chiral superfield
\beq
\label{CA6}
{\cal W} = -{i\over2}\, \ov{\widetilde D\widetilde D} \, \widehat{\cal W}
+ {i\over2}\, \ov{DD} \, \ov{\widehat{\cal W}},
\eeq
one finds
\beq
\label{CA7}
\begin{array}{rcl}
{\cal W} &=& \displaystyle  X 
+ \sqrt2i\, \widetilde\theta^\alpha \ov D_\dalpha \Bigl[ D_\alpha\ov\Omega^\dalpha 
+ {1\over2}\ov D^\dalpha\Omega_\alpha \Bigr]
- \widetilde\theta\widetilde\theta\, {1\over4} \, \ov{DD} \, \ov X .
\crbig
&=& 
\displaystyle X 
+ \sqrt2i\, \widetilde\theta^\alpha W_\alpha
- \widetilde\theta\widetilde\theta\, {1\over4} \, \ov{DD} \, \ov X ,
\end{array}
\eeq
where $W_\alpha$ is the usual Maxwell chiral superfield 
$$
W_\alpha = -{1\over4} \ov{DD} D_\alpha V
$$
with, however, 
\beq
\label{CA8}
V = 2 ( \mathbb{L} + \ov{\mathbb{L} })
\eeq
instead of $V$ being simply a real superfield. This new condition follows from
\beq
\label{CA9}
\ov D_\dalpha \Bigl[ D_\alpha\ov\Omega^\dalpha 
+ {1\over2}\ov D^\dalpha\Omega_\alpha \Bigr] = -{1\over2}\ov{DD} D_\alpha (\mathbb{L}+\ov\mathbb{L})\,,
\eeq
which is a consequence of (\ref{CA6}). The ${\cal N}=2$ gauge transformation of $\widehat{\cal W}$ leaving ${\cal W}$ invariant can be read from 
expressions
(\ref{CA7}) and (\ref{CA8}): ${\cal W}=0$ if $X=0$ and $\mathbb{L} = iL$, with a real linear $L$. In other words,
${\cal W}$ is invariant under 
\beq
\label{CA10}
\widehat{\cal W} \qquad\longrightarrow\qquad \widehat{\cal W} +  {\cal Y},
\qquad\qquad
{\cal Y} = U + \sqrt2i\, \ov{\widetilde\theta \, D } L - \ov{\widetilde\theta\widetilde\theta}\,{1\over4}\ov{DD}\,\ov U.
\eeq
Eq.~(\ref{longM0}) indicates that this gauge variation is induced by a single--tensor supermultiplet in a ``short"
chiral--antichiral superfield. 

\subsection{The long and short super--Maxwell superfields}

To summarize, to describe the single--tensor and the Maxwell multiplet, we have obtained two pairs of ${\cal N}=2$ superfields respectively, with each pair containing one long ($16_B+16_F$) and one
short ($8_B+8_F$) superfield:
\begin{center} 
\begin{tabular}{ llll } \hline \vspace{-2mm} \\
& Long, $16_B+16_F$ \quad & Short, $8_B+8_F$ \quad & Gauge variation, $8_B+8_F$
\crbig \hline \\
Maxwell: & $\widehat{\cal W}$ & ${\cal W}$ & $\delta\,\widehat{\cal W} = {\cal Z}_{gauge} \qquad
\delta\,{\cal W} =0$
\crbig 
Single--tensor: \quad & $\widehat{\cal Z}$ & ${\cal Z}$ &  $\delta\,\widehat{\cal Z} = {\cal W}_{gauge} \qquad 
\delta\,{\cal Z} =0$ 
\\ \vspace{-2mm}
\\ \hline 
\end{tabular} 
\end{center}
Counting off--shell degrees of freedom in the ``long" Maxwell multiplet is interesting. Firstly, $X$ and $U$ include
$8_B+8_F$ fields while the complex linear $\mathbb{L}$ has $12_B+12_F$ components.\footnote{It is a complex
superfield $(16_B+16_F)$ with the chiral constraint $\ov{DD}\,\mathbb{L}=0$, removing $4_B+4_F$ fields.} 
The superfield
$\widehat{\cal W}$ depends however on $\ov D_\dalpha \, \mathbb{L}$ and one can write 
$\mathbb{L} = \Phi + \Delta\mathbb{L}$ ($\Phi$ chiral), with $8_B+8_F$ fields in $\Delta\mathbb{L}$: the superfield
$\widehat{\cal W}$ sees then only $16_B+16_F$ fields. One actually expects that a larger supermultiplet with
$24_B+24_F$ fields exists, with all ${\cal N}=2$ partners of $\mathbb{L}$. This is discussed in Appendix
\ref{AppB}.

The variation (\ref{CA10}) is not the gauge transformation of the super--Maxwell theory: it does not act on 
$V= 2 ( \mathbb{L} + \ov{\mathbb{L} })$. It only allows to eliminate $U$ and $4_B+4_F$ components of $\mathbb{L}$, leaving $X$, $V$, $W_\alpha$
and then also the ${\cal N}=2$ superfield ${\cal W}$ unchanged. 
The standard Maxwell gauge transformation $V \longrightarrow V+ \Lambda+ \ov\Lambda$ is actually
\beq
\label{CA11}
\mathbb{L} \qquad\longrightarrow\qquad \mathbb{L} + {1\over2}\,\Lambda, \qquad\qquad \ov D_\dalpha\,\Lambda=0,
\eeq
which is a symmetry of $\widehat{\cal W}$.\footnote{See Appendix \ref{AppB}.} A comparison of $2(\mathbb{L}+\ov\mathbb{L})$ with the standard 
expansion of the Maxwell real superfield indicates that the gauge field and the auxiliary $\theta\theta\ov{\theta\theta}$ component are respectively
\beq
\label{CA12}
\begin{array}{rcl}
A_\mu &=& -4\Re \mathbb{V}_\mu\,,
\crbig
D &=& -4 \, \partial^\mu \Im\mathbb{V}_\mu \,.
\end{array}
\eeq
Replacing the scalar $D$ by the divergence of a vector field has nontrivial consequences which
are precisely discussed in Appendix \ref{AppC}. In short, the role of the FI coefficient $\xi$ is taken
by an integration constant appearing when solving the field equation of $\Im \mathbb{V}_\mu$
and a well--defined procedure for the elimination of $\Im \mathbb{V}_\mu$ shows that the theories
formulated with either $D$ or $\Im \mathbb{V}_\mu$ are physically equivalent. 

\subsection{Long superfield and nonlinear deformations}

According to relation (\ref{CA6}), the nonlinear deformation ${\cal W}_{nl}$ can be transferred to a deformation $\widehat{\cal W}_{nl}$ only
if $A^2=\ov B^2$, $\Gamma=0$ since the only available chiral--antichiral deformation term would be
\beq
\widehat{\cal W}_{nl} = - {i\over2} \, A^2 \, \theta\theta\,\ov{\widehat\theta\widehat\theta}.
\eeq
 This is the case if the deformation can be viewed as a background value of the
auxiliary $F$ in $X$, which never leads to partial breaking. A similar argument holds for the single--tensor 
superfield with relation (\ref{longM2}). Then, to consider a general deformation and in particular if the interest 
is in partial supersymmetry breaking, the deformed short version of the superfields must be used. 
Since these short superfields have different chiralities, writing an interaction of two deformed 
supermultiplets is problematic.

\section{Interactions}

\subsection{The Chern-Simons interaction}
\setcounter{equation}{0}

The interaction of a ${\cal N}=2$ Maxwell multiplet with a single--tensor multiplet can be introduced either
by a supersymmetrization of the Chern--Simons coupling $B\wedge F$ or by a supersymmetrization 
of $F_{\mu\nu}-B_{\mu\nu}$. These options are related via electric--magnetic duality. 
The supersymmetric interaction exists for off--shell fields and can be written in ${\cal N}=2$ or ${\cal N}=1$ 
superspace. 
The goal of this Subsection is to discuss the Chern--Simons coupling of a nonlinear or constrained Maxwell or
single--tensor multiplet with unbroken linear ${\cal N}=1$, to its counterpart with linear ${\cal N}=2$.

In terms of ${\cal N}=1$ superfields, the ${\cal N}=2$ Chern--Simons interaction
can be written in two simple ways.
Firstly, using $(L,\Phi)$ and $(V_1,V_2)$ to describe the single--tensor and Maxwell multiplets respectively, 
the Chern--Simons interaction with (real) coupling $g$ can be written as a ${\cal N}=1$ $D$--term \cite{ADM}, \cite{AADT}:
\beq
\label{CS1}
{\cal L}_{CS} = - g\Dint \Bigl[ V_1(\Phi+\ov\Phi) + V_2L \Bigr]\,.
\eeq
It is invariant under the second supersymmetry variations (\ref{part3}) and (\ref{B1}) and it is also gauge invariant.
A second expression using an $F$--term exists in terms of $\chi_\alpha$, $\Phi$ for the single--tensor and 
$X$, $W_\alpha$ for the Maxwell multiplet, using the relations
$$
L = D\chi-\ov{D\chi}, \qquad W_\alpha= -{1\over4}\ov{DD}D_\alpha V_2, \qquad X={1\over2}\ov{DD}V_1
$$
and some partial integrations:
\beq
\label{CS2}
{\cal L}_{CS} = g\Fint \Bigl[ {1\over2}\Phi X + \chi^\alpha W_\alpha \Bigr] 
+ g\Fbarint \Bigl[ {1\over2}\ov\Phi \ov X - \ov\chi_\dalpha \ov W^\dalpha \Bigr] .
\eeq
The expressions (\ref{CS1}) and ({\ref{CS2}}) differ by a derivative term. The chiral form can be extended to a chiral integral
over ${\cal N}=2$ superspace, using the chiral--chiral superfields ${\cal W}$ and $\widehat{\cal Z}$
for the Maxwell and single--tensor multiplets respectively \cite{AADT}:\footnote{See eqs.~(\ref{Max1}) and (\ref{longM1}).} 
\beq
\label{CS3}
{\cal L}_{CS} = ig \Fint\bigint d^2\widetilde\theta\, {\cal W} \widehat{\cal Z} + {\rm h.c.}
\eeq
All dependence on $Y$ disappears in the imaginary part of 
$[ {\cal W} \widehat{\cal Z} ]_{\widetilde\theta\widetilde\theta}$ (under a spacetime integral). This expression is also invariant under the gauge transformation (\ref{longM5}) of $\widehat{\cal Z}$, since, for any pair of (short) Maxwell multiplets ${\cal W}_1$ 
and ${\cal W}_2$, 
$$
\Im \Fint \bigint d^2\widetilde\theta\, {\cal W}_1{\cal W}_2 \qquad\quad{\rm and}\qquad\quad
\Im \Fint W_1^\alpha W_{2\alpha} 
$$
are derivative terms.

Finally, one can also write the Chern--Simons lagrangian using the chiral--antichiral superfields ${\cal Z}$
(short) and $\widehat{\cal W}$ (long) for the single--tensor and the Maxwell multiplet respectively\footnote{Eqs.~(\ref{st2}) 
and (\ref{CA5}).}
\beq
\label{CS4}
{\cal L}_{CS} = i g\Fint\bigint d^2\ov{\widetilde\theta}\, \widehat{\cal W}{\cal Z} + {\rm h.c.}
\eeq
This can be verified either by direct calculation or by using relation (\ref{CA6}) and partial integrations in
expression (\ref{CS3}) and of course $V_2 = 2(\mathbb{L} + \ov\mathbb{L})$. Equation (\ref{CS4}) is invariant up to a derivative term under the gauge transformation (\ref{B13}) of $\widehat{\cal W}$, since, for any pair of (short) single--tensor multiplets ${\cal Z}_1$, ${\cal Z}_2$, 
$$
\Im \Fint \bigint d^2\ov{\widetilde\theta}\, {\cal Z}_1{\cal Z}_2 \qquad\quad{\rm and}\qquad\quad
\Im \Fint (\ov{D}_{\dot{\alpha}}L_1) (\ov{D}^{\dot{\alpha}}L_2) 
$$
are derivative terms.

In terms of the ${\cal N}=1$ component superfields,
\beq
\label{CS5}
{\cal L}_{CS} =g \Fint \Bigl[ {1\over2}\Phi X + (\ov DL)(\ov D\mathbb{L}) \Bigr] 
+ g\Fbarint \Bigl[ {1\over2}\ov\Phi \ov X + (DL)(D\ov\mathbb{L}) \Bigr] .
\eeq
In components, using expansions (\ref{part6b}) and (\ref{CA4}), we find that (under a spacetime integral)
\beq
\label{CS6}
\begin{array}{rcl}
{\cal L}_{CS} &=& - \frac{1}{2} g (xf+ \ov{x} \ov{f} +zF+\ov{z}\ov{F} + \kappa \psi + \ov{\kappa}\ov{\psi}) + ig\lambda\varphi - ig\ov{\lambda\varphi} 
\crbig
&&-{1\over8}g\, \epsilon_{\mu\nu\rho\sigma} B^{\mu\nu} F^{\rho\sigma}
+ 2 g\, C\partial_\mu \Im \mathbb{V}^\mu - g\,\partial_\mu\varphi\sigma^\mu\ov\omega -g\,\omega\sigma^\mu\partial_\mu\ov\varphi
\,,
\end{array}
\eeq
where $F^{\rho\sigma} \equiv  \partial^\rho A^\sigma - \partial^\sigma A^\rho \,$.

\subsubsection{The Chern-Simons interaction with deformed Maxwell multiplet} \label{secb}

The nonlinearly--deformed Maxwell multiplet is described by the CC superfield ${\cal W}$, including the deformation terms
(\ref{Yis3}). This leads to the Chern--Simons interaction
\beq
\label{CS7}
\begin{array}{rcl}
{\cal L}_{nl} &=& ig\, \Fint\bigint d^2\widetilde\theta\,  \widehat{\cal Z}\,{\cal W}  + {\rm h.c.}
\crbig
&=& \displaystyle {\cal L}_{CS} + i g\,\Fint\left[ B^2 Y - \sqrt2\,\Gamma \, \theta\chi 
- A^2\,\theta\theta \left({i\over2}\Phi + {1\over4}\ov{DD}\ov Y \right) \right] + {\rm h.c.}\,,
\end{array}
\eeq
where ${\cal L}_{CS} $ is given by (\ref{CS2}). For the partial breaking, using $A=\Gamma=0$, we obtain
\beq
\label{CS8}
{\cal L}_{nl} = g\Fint \Bigl[ {1\over2}\Phi X + \chi^\alpha W_\alpha + iB^2 Y \Bigr] + {\rm h.c.} 
\eeq
The second supersymmetry variation $\sqrt2iB^2 \eta\chi$ of $iB^2Y$ is cancelled by the nonlinear
variation of $W_\alpha$, $\delta^*W_\alpha = -\sqrt2 iB^2 \eta_\alpha + {\rm linear}$. However, the equation of motion of $Y$ is inconsistent. One can get around this problem by using $l > 1$ deformed Maxwell multiplets (namely one ``long'' single--tensor and at least two ``short'' and deformed Maxwell multiplets), as then the relevant equation of motion would take the form of a tadpole--like condition
\beq \label{tadpole}
g_a B^2_a=0 \quad , \quad a=1,...,l \,,
\eeq
where $g_a$ would be the coupling of each Chern--Simons interaction. This is in agreement with the claim made in \cite{PartPi} and \cite{Fujiwara:2005hj}, namely that one cannot couple hypermultiplets to a single Maxwell multiplet in a theory with partial breaking induced by the latter.

The Chern--Simons interaction (\ref{CS8}) can be combined with the kinetic lagrangian
\beq
\label{CS9}
{\cal L}_{kin.} = \Dint {\cal H}(L, \Phi, \ov\Phi) + \frac{1}{2} \Fint\bigint d^2\widetilde\theta\, {\cal F}({\cal W} ) 
+{\rm h.c.}
\eeq
for the two multiplets, as well as with an FI contribution
\beq
\label{CS10}
{\cal L}_{FI} = \xi\Dint V_2 +\frac{1}{2} m^2 \Fint X + {\rm h.c.}
\eeq
The theory depends then on a function ${\cal H}$ solving the Laplace equation and on an arbitrary holomorphic function ${\cal F}$. Imposing the constraint ${\cal W}^2=0$ (where ${\cal W}$ is deformed) eliminates $X$, which becomes a function $X(WW)$ of $WW$ and its derivatives. Moreover, due to the constraint, the lagrangian no longer depends on ${\cal F}$ and it reduces to 
\beq
\Dint {\cal H}(L, \Phi, \ov\Phi) +\xi\Dint V_2 +\frac{1}{2} m^2 \Fint X + {\rm h.c.}
\eeq 
The resulting theory has a linear ${\cal N}=1$ as well as a second nonlinear supersymmetry and has been analyzed in~\cite{AADT}. 

\subsubsection{The Chern-Simons interaction with deformed single--tensor multiplet} 

In the analogous procedure for the nonlinear single--tensor multiplet, the CA superfield (\ref{st2}) with deformation
(\ref{st4}) is coupled to the long Maxwell CA superfield (\ref{CA5}):
\beq
\label{CS11A}
\begin{array}{rcl}
{\cal L}_{nl} &=& ig\, \Fint\bigint d^2\ov{\widetilde\theta}\,  \widehat{\cal W} \, {\cal Z} + {\rm h.c.}
\crbig
&=& \displaystyle {\cal L}_{CS} + ig\, \Fint\left[ \ov{\widetilde{B}}^2 U 
- \widetilde{A}^2\,\theta\theta \left({i\over2}X + {1\over4}\ov{DD}\ov U \right) \right] + {\rm h.c.}\,,
\end{array}
\eeq
where ${\cal L}_{CS}$ is given by (\ref{CS5}). Requiring now partial breaking with $\widetilde{A}=0$ yields
\beq
\label{CS12}
{\cal L}_{nl} = g\Fint \Bigl[ {1\over2}\Phi X + (\ov DL)(\ov D\mathbb{L}) + i\ov{\widetilde{B}}^2 U \Bigr] + {\rm h.c.} 
\eeq
Since\footnote{See Appendix \ref{AppB}.}
$$
\delta^*\, i\ov{\widetilde{B}}^2 U = \sqrt2i \, \ov{\widetilde{B}}^2 \, \ov{\eta D} \, \mathbb{L} \,, \qquad\qquad
\delta^* \ov D_\dalpha L = - \sqrt 2i \, \ov{\widetilde{B}}^2 \, \ov\eta_\dalpha \,,
$$
${\cal L}_{nl}$ is invariant under a linear ${\cal N}=1$ and under a second nonlinear supersymmetry. However, the equation of motion of $U$ is inconsistent as that of $Y$ of the previous Subsection -- this problem can be solved by  coupling the ``long'' Maxwell multiplet(s) to at least two ``short'' and deformed single--tensor multiplets\footnote{Note that there is no reason to identify the imaginary part of the auxiliary field of $U$ with a four--form field as was done for $Y$ in \cite{AADT}. In particular, the variation of $Y$ under the gauge transformation of $\widehat{\cal Z}$ is $\delta_{gauge}Y=-\frac{1}{2}\ov{D}\ov{D}\Delta'$ \cite{AADT}, where $\Delta'$ is a real superfield, while the variation of $U$ under the gauge transformation of $\widehat{\cal W}$ is $\delta_{gauge}U=\Sigma_c$ (see (\ref{B12}) of Appendix \ref{AppB}) and the chiral superfield $\Sigma_c$ is not necessarily identified with $\ov{D}\ov{D}\Delta''$, where $\Delta''$ is a real superfield.}.

The complete theory has then lagrangian
\beq
\label{CS11f}
\begin{array}{rcl}
{\cal L} &=& {\cal L}_{nl} + \Bigg[ \frac{1}{2} \Fint\bigint d^2\widetilde\theta  {\cal F}({\cal W})
+  \Fint \bigint d^2\ov{\widetilde\theta}\,{\cal G}({\cal Z}) + \Fint \widetilde{m}^2\Phi \Bigg] + {\rm h.c.} 
\crbig
&& +\xi\Dint V_2\,,
\end{array}
\eeq
where ${\cal Z}$ is deformed and we have added an FI term for $V_2$.  Upon imposing the constraint (\ref{st14}), ${\cal G}$ does not contribute to (\ref{CS11f}), since
\beq
 \Fint \bigint d^2\ov{\widetilde\theta}\, {\cal Z} +{\rm h.c.} \sim \Fint \ov{D}^2 \ov{\Phi}+{\rm h.c.} = \textrm{deriv. term}
\eeq
and the bosonic part of (\ref{CS11f}) becomes
\beq
\label{CSd}
\begin{array}{rcl}
{\cal L}_{bos}&=&\frac{1}{2} \Fint\bigint d^2\widetilde\theta \, {\cal F}({\cal W})|_{bos}+{\rm h.c.} -2 \xi \,\partial^\mu \Im \mathbb{V}_\mu 
\crbig
&& +2g\, \Big(- \frac{1}{26} \epsilon_{\mu \nu \rho \sigma} H^{ \nu \rho \sigma} A^{\mu}+  C \,\partial^\mu \Im \mathbb{V}_\mu - \widetilde{B}^2 \Im F_U\Big) 
\crbig
&& + (g \Re x +2 \widetilde{m}^2) \widetilde{B}^2
\crbig 
&& \hspace{1.2cm} \cdot
\Bigg(1- \sqrt{1- \frac{2}{\widetilde{B}^4} \Big(\frac{1}{6}H_{\mu \nu \rho} H^{\mu \nu \rho} + \partial_\mu C \partial^\mu C \Big)  - \frac{1}{9\widetilde{B}^8}(\epsilon_{\mu \nu \rho \sigma} H^{ \nu \rho \sigma} \partial^{\mu} C)^2} \,\Bigg) \,,
\end{array}
\eeq
where $\widetilde{B}$ has been assumed to be real and $F_U$ is the auxiliary field of $U$. Notice that the lagrangian (\ref{dbian}) has acquired a field--dependent coefficient $ (g \Re x +2 \widetilde{m}^2) \widetilde{B}^2$ as its analogue, the Born--Infeld lagrangian, does in ref. \cite{AADT}.

The solution of the equation of motion for the auxiliary field $F$ of $X$ is $F=0$. Moreover, the equation of motion for the auxiliary field $\Im \mathbb{V}_\mu$ is
\beq
\partial_\mu \big(16 \Re \mathcal{F}_{xx}\, \partial^\nu \Im \mathbb{V}_\nu +2g\,C\big) =0\,,
\eeq
whose solution is
\beq
16 \Re \mathcal{F}_{xx}\, \partial^\nu \Im \mathbb{V}_\nu +2g\,C = -\lambda\,,
\eeq
where $\lambda$ is an arbitrary integration constant. For reasons explained in Appendix C, we make the identification
\beq
\lambda = 2\xi\,.
\eeq
The scalar potential of the theory is then
\beq
V= \frac{1}{32 \Re \mathcal{F}_{xx}} (2g\,C -2\xi)^2\,,
\eeq
whose supersymmetric vacuum is at
\beq
<C> =\frac{\xi}{g}\,.
\eeq

In this vacuum, $x$ corresponds to a flat direction of the potential and is massless. The canonically normalized mass $\mathcal{M}^2_{C,can}$ that $C$ aquires is then
\beq
\mathcal{M}^2_{C,can} = \frac{1}{4}\frac{1}{\Re \mathcal{F}_{xx}} \,\frac{g^2 \widetilde{B}^2}{2g\, \Re x +4 \widetilde{m}^2}\,. 
\eeq
Moreover, the interaction term $- \frac{1}{12} g\,\epsilon_{\mu \nu \rho \sigma} H^{ \nu \rho \sigma} A^{\mu}$ generates a mass term for $A_\mu$ and we find that the canonically normalized mass $\mathcal{M}^2_{A_\mu,can}$ is
\beq
\mathcal{M}^2_{A_\mu,can} = \mathcal{M}^2_{C,can} \,.
\eeq
The spectrum consists then of a massive ${\cal N}=1$ vector multiplet and a massless ${\cal N}=1$ chiral multiplet $X$; the Chern--Simons coupling results in the vector multiplet $W$ absorbing the goldstino multiplet, while $X$ remains massless. Consequently, we observe a mechanism analogous to the super--Brout--Englert--Higgs effect without gravity \cite{AADT}, which is induced by the Chern--Simons coupling of the previous Subsection (\ref{secb}).

\subsection{Constrained matter multiplets}

In Subsection 6.1, we described the couplings of the deformed ${\cal N}=2$ goldstino multiplet to unconstrained matter ${\cal N}=2$ multiplets. They are based on a Chern--Simons interaction that couples a Maxwell to a single--tensor multiplet, where one of the two contains the goldstino. In both cases, upon imposing a nilpotent constraint on the goldstino multiplet, the Chern--Simons interaction generates a super--Brout--Englert--Higgs phenomenon without gravity, where the goldstino is absorbed in a massive ${\cal N}=1$ vector multiplet, while a massless chiral multiplet remains in the spectrum.

Here, we discuss generalisations of the nilpotent constraint in order to describe, besides the goldstino, incomplete matter multiplets of non--linear supersymmetry in which half of the degrees of freedom are integrated out of the spectrum, giving rise to constraints. Examples of such constraints in ${\cal N}=1$ non--linear supersymmetry, which is described by the nilpotent goldstino superfield $X$ with $X^2=0$, are given by
\beq \label{ferm}
X\Phi=0\,,
\eeq which eliminates the scalar component of the matter chiral superfield $\Phi$, or
\beq \label{scalar}
 X\ov\Phi= \textrm{chiral}\,,
\eeq
that eliminates the fermion component of $\Phi$ \cite{komar}. In ${\cal N}=2$, we examine below both cases, with the goldstino being part of either a nilpotent (deformed) Maxwell multiplet $\cal W$ with ${\cal W}^2=0$, or of a nilpotent (deformed) single--tensor multiplet $\cal Z$ with ${\cal Z}^2=0$.

\subsubsection{The goldstino in the Maxwell multiplet}

Consider the case in which the goldstino is in a deformed Maxwell multiplet $\mathcal{W}_0$, given by (\ref{firstw})
\beq 
{\cal W}_0 = X_0 + \sqrt2i\,\widetilde\theta W_0 + \widetilde\theta\widetilde\theta \, \left[ B^2
- {1\over4}\ov{DD}\,\ov X_0 \right] \,,
\eeq
which satisfies the constraint ${\cal W}_0^2=0$,
or, equivalently, eq.~(\ref{Mcon3}) \cite{BG}: 
\beq \label{qep1}
X_0 = -2 \, \frac{W_0 W_0}{4B^2-\ov{D}\ov{D}\ov{X}_0}\,.
\eeq
To describe an incomplete ${\cal N}=2$ vector multiplet with non--linear supersymmetry containing an ${\cal N}=1$ vector $W_1$, we consider the ${\cal N}=2$ constraint
\beq \label{sh1}
{\cal W}_0 {\cal W}_1 =0\,, 
\eeq
where $ {\cal W}_1 $ is an undeformed (and short) Maxwell multiplet given by (\ref{Max1}):
\beq
{\cal W}_1 = X_1 + \sqrt2i\,\widetilde\theta W_1 - {1\over4}\widetilde\theta\widetilde\theta \,\ov{DD}\,\ov X_1 \,.
\eeq
The constraint (\ref{sh1}) then yields the following set of equations
\beq \label{syst1}
\begin{array}{ccc}
X_0 X_1 &=&0\,,
\crbig
X_0W_{1\alpha}+ X_1 W_{0\alpha}&=&0\,,
\crbig
X_1 B^2-\frac{1}{4}\ov D \ov D (X_0 \ov{X}_1+X_1 \ov{X}_0) + W_0 W_1 &=&0\,.
\end{array}
\eeq

We now use (\ref{qep1}) and the identity
\beq
(W_0 W_1) W_{0\alpha} = - \frac{1}{2} (W_0 W_0) W_{1 \alpha}
\eeq
to solve the second of equations (\ref{syst1}), which yields
\beq \label{sh2}
X_1 = -4\,\frac{W_0 W_1}{4B^2-\ov{D}\ov{D}\ov{X}_0} + h\, W_0 W_0\,,
\eeq
where $h$ is a chiral superfield. This expression verifies the first eq.~(\ref{syst1}) for all $h$
and the third eq.~(\ref{syst1}) if
\beq
h = -2\, \frac{\ov{D}\ov{D}\ov{X}_1}{(4B^2-\ov{D}\ov{D}\ov{X}_0)^2}
\eeq
and thus
\beq \label{shh}
X_1 = -4\,\frac{W_0 W_1}{4B^2-\ov{D}\ov{D}\ov{X}_0} -2\, \frac{\ov{D}\ov{D}\ov{X}_1}{(4B^2-\ov{D}\ov{D}\ov{X}_0)^2}\, W_0 W_0\,.
\eeq
One may further use the solution (\ref{opk}) for $X_0$ and solve (\ref{shh}) to obtain $X_1$ as a function of $W_0$, $W_1$ and their derivatives; the constraint (\ref{sh1}) eliminates $X_1$.

Note that the constraint ${\cal W}_0^2={\cal W}_0{\cal W}_1=0$ is a particular case of the system of equations
\beq \label{dijk}
d_{abc}{\cal W}_b{\cal W}_c=0\quad;\quad a,b,c=1,\dots,l
\eeq
introduced in~\cite{Ferrara:2014oka, Ferrara:2014oka2} to obtain coupled DBI (Dirac--Born--Infeld) actions. In eqs.~(\ref{dijk}), all ${\cal W}_a$ are in general deformed with different deformation parameters $B_a$ and the constants $d_{abc}$ are totally symmetric. Our constraints correspond to the case of two ${\cal N}=2$ vector multiplets with $d_{000}=d_{001}=1$ and all other $d$'s vanishing.

We can also describe incomplete ${\cal N}=2$ single--tensor multiplets containing a single ${\cal N}=1$ chiral multiplet. For that, let us consider the constraint
\beq \label{sh3}
{\cal W}_0 \widehat{\cal Z}=0\,, 
\eeq
where $\widehat{\cal Z}$ is a ``long''\footnote{Note that it is easy to check that the constraint ${\cal W}_0 {\cal Z}=0$, where ${\cal Z}$ is a ``short'' single--tensor multiplet, leads to an overconstrained system of equations.} single--tensor multiplet given by (\ref{longM1}). Equation (\ref{sh3}) then leads to
\beq \label{syst2}
\begin{array}{ccc}
X_0 Y &=&0\,,
\crbig
X_0\chi_{\alpha}+ iY W_{0\alpha}&=&0\,,
\crbig
Y B^2 - \frac{i}{2} \, \Phi X_0 -\frac{1}{4}\ov D \ov D (X_0 \ov{Y}+Y \ov{X}_0) - i \, W_0 \chi &=&0\,,
\end{array}
\eeq
which, following the same steps as before, yield
\beq \label{sh4}
Y = 4i\,\frac{W_0 \chi}{4B^2-\ov{D}\ov{D}\ov{X}_0} -2\, \frac{2i\Phi+\ov{D}\ov{D}\ov{Y}}{(4B^2-\ov{D}\ov{D}\ov{X}_0)^2}\, W_0 W_0\,,
\eeq
which again one may solve to eliminate $Y=Y(W_0,\chi,\Phi)$. 

One can also check if the expression (\ref{sh4}) is covariant under the gauge variation (\ref{longM5})
\beq \label{sh6}
\widehat{\cal Z} \qquad\longrightarrow\qquad \widehat{\cal Z} + {\cal W}_g\,,
\eeq
where ${\cal W}_g$ is a ``short'' (undeformed) Maxwell multiplet with components $(X_g,  W_{g\alpha})$, or, equivalently,
\beq \label{sh5}
\delta Y = X_g \quad , \quad \delta\chi_{\alpha} = i W_{g\alpha} \quad , \quad \delta \Phi =0\,.
\eeq
Under (\ref{sh5}), the expression (\ref{sh4}) becomes
\beq \label{sh11}
X_g= -4\,\frac{W_0 W_g}{4B^2-\ov{D}\ov{D}\ov{X}_0} -2\, \frac{\ov{D}\ov{D}\ov{X}_g}{(4B^2-\ov{D}\ov{D}\ov{X}_0)^2}\, W_0 W_0\,,
\eeq
which, as was previously shown, is actually the consequence of
\beq \label{sh12}
{\cal W}_0 {\cal W}_g =0\,, 
\eeq
that is the variation of (\ref{sh3}) under (\ref{sh6}). The expression (\ref{sh4}) is thus invariant 
only under the reduced gauge transformations (\ref{sh6}) subject to the constraint (\ref{sh12}). These are not sufficient to eliminate all unphysical components of $\widehat{\cal Z}$. 

Alternatively, we can consider that we actually solve the constraints ${\cal W}_0 (\widehat{\cal Z} - {\cal W}_g)
={\cal W}_0{\cal W}_g=0$, where $\widehat{\cal Z} - {\cal W}_g$ is gauge invariant and ${\cal W}_g$ can be 
eliminated by a gauge transformation (\ref{sh6}).
One can then choose $Y- X_g=0$ and use eq.~(\ref{sh4}) to eliminate $\chi - iW_g$ in terms of the 
${\cal N}=1$ chiral superfield $\Phi$:
\beq \label{sh13}
\chi_\alpha - iW_g{}_\alpha = \frac{\Phi}{(4B^2-\ov{D}\ov{D}\ov{X}_0)}\, W_{0\alpha}\,.
\eeq
In the physically--relevant linear superfield $L$ however, $W_g$ disappears:
$$
L = D\chi-\ov{D\chi} = D(\chi-iW_g) -\ov D(\ov\chi-i\ov W_g) \,,
$$
since $W_g$ verifies the Bianchi identity.

\subsubsection{The goldstino in the single--tensor multiplet}

Now let us consider the case in which the goldstino is in a deformed single--tensor multiplet ${\cal Z}_0$, given by
\beq 
{\cal Z}_0 = \Phi_0 + \sqrt2i \, \ov{\widetilde\theta D} L_0 + \ov{\widetilde\theta\widetilde\theta}\, \left[ \ov{\widetilde{B}}^2
- {1\over4}\,\ov{DD}\,\ov\Phi_0 \right] \,,
\eeq
which satisfies (\ref{st14})
\beq
{\cal Z}_0^2=0\,,
\eeq
or equivalently eq.~(\ref{st12}) \cite{Bagger:1997pi}: 
\beq
\Phi_0 = -2\, {(\ov DL_0)(\ov DL_0) \over 4\ov{\widetilde{B}}^2 - \ov{DD}\ov\Phi_0}  \,.
\eeq

To describe another incomplete ${\cal N}=2$ single--tensor multiplet with non--linear supersymmetry containing an ${\cal N}=1$ linear multiplet, we consider the ${\cal N}=2$ constraint
\beq
{\cal Z}_0 {\cal Z}_1 =0\,,
\eeq
where ${\cal Z}_1$ is an undeformed (and short) single--tensor multiplet given by (\ref{st2})
\beq
{\cal Z}_1 = \Phi_1 + \sqrt2i \, \ov{\widetilde\theta D}L_1 - {1\over4}\,\ov{\widetilde\theta\widetilde\theta}\, \ov{DD}\,\ov\Phi_1\,.
\eeq
Following the same steps as before, as well as the identity
\beq
(\ov{ D}L_0\ov{ D}L_1)\ov{D}_{\dot{\alpha}}L_0 = - \frac{1}{2} \,(\ov{ D}L_0\ov{ D}L_0)\ov{D}_{\dot{\alpha}}L_1\,,
\eeq
we find
\beq \label{sh10}
\Phi_1 = -4 \, \frac{\ov{ D}L_0\ov{ D}L_1}{4\ov{\widetilde{B}}^2 - \ov{DD}\ov\Phi_0}-2\, \frac{\ov{DD}\ov\Phi_1}{(4\ov{\widetilde{B}}^2 - \ov{DD}\ov\Phi_0)^2}\ov{ D}L_0\ov{ D}L_0\,,
\eeq
which one may solve to eliminate the chiral component $\Phi_1$ in terms of $L_1$ and the goldstino multiplet $L_0$. Note that the constraints ${\cal Z}_0^2={\cal Z}_0 {\cal Z}_1=0$ can be generalised to a system of equations
\beq \label{hijk}
\widetilde{d}_{abc}{\cal Z}_b{\cal Z}_c=0\quad;\quad a,b,c=1,\dots,l\,,
\eeq
in analogy with the system (\ref{dijk}), where $\widetilde{d}_{abc}$ are totally symmetric constants, in order to obtain a coupled action of non--linear (deformed) single--tensor multiplets.

Finally, we consider the constraint
\beq \label{sh7}
{\cal Z}_0 \widehat{\cal W}=0 \,,
\eeq
where $ \widehat{\cal W}=0$ is a ``long'' Maxwell multiplet given by (\ref{CA1}), and, using the same procedure as before, we obtain
\beq \label{sh8}
U = 4i \,  \frac{\ov{ D}L_0\ov{ D}\mathbb{L}}{4\ov{\widetilde{B}}^2 - \ov{DD}\ov\Phi_0}-2\, \frac{2iX+\ov{DD}\ov U}{(4\ov{\widetilde{B}}^2 - \ov{DD}\ov\Phi_0)^2}\ov{ D}L_0\ov{ D}L_0\,,
\eeq
which eliminates $U$. Using the same reasoning as before, one can show that the solution (\ref{sh8}) is invariant under the reduced gauge variation (\ref{CA10}) 
\beq
\label{shgauge}
\widehat{\cal W}\qquad\longrightarrow\qquad \widehat{\cal W} + {\cal Z}_g\,,
\eeq
where ${\cal Z}_g$ is a ``short'' (undeformed) single--tensor multiplet, namely 
$\delta U = \Phi_g\,,\,\delta \mathbb{L} = i L_g\,,\,\delta X=0$,
satisfying the constraint
\beq\label{sh20}
{\cal Z}_0 {\cal Z}_g=0\, .
\eeq
Following the same procedure as for the solution of the constraint (\ref{sh3}), one can use the full gauge invariance to set $U=0$. Eq.~(\ref{sh8}) can then be used to eliminate $\ov\Omega_\dalpha = \ov D_\dalpha\,\mathbb{L}$ in terms of the ${\cal N}=1$ chiral superfield $X$:\footnote{Since $L$ is real linear, $SL$ is complex linear for any chiral $S$.}
\beq \label{sh9}
\ov{ D}_\dalpha\,\mathbb{L}= \frac{X}{4\ov{\widetilde{B}}^2 - \ov{DD}\ov\Phi_0}\ov{ D}_\dalpha\,L_0\,.
\eeq
This result defines $\mathbb{L}$ up to the addition of an arbitrary chiral field:
as expected, the constraint equation (\ref{sh9}) is invariant under  the Maxwell gauge transformation
$$
\mathbb{L} \qquad\longrightarrow\qquad \mathbb{L} + \Lambda_c, \qquad\qquad \ov D_\dalpha\Lambda_c = 0
$$
(see Appendix \ref{AppB}). In addition, the physically--relevant $V=2(\mathbb{L}+\ov\mathbb{L})$ in invariant 
under the gauge ambiguity (\ref{shgauge}).

\section{Conclusions}
\setcounter{equation}{0}

In this work, we studied the off--shell partial breaking of global ${\cal N}=2$ supersymmetry using constrained ${\cal N}=2$ superfields. The corresponding Goldstone fermion belongs to a vector or a linear multiplet of the unbroken ${\cal N}=1$ supersymmetry and is described by a deformed ${\cal N}=2$ Maxwell or single--tensor superfield, respectively, satisfying a nilpotent constraint. Unlike ${\cal N}=1$ non--linear supersymmetry, where the nilpotent constraint assumes a non--vanishing expectation value for the F--component of the goldstino superfield arising a priori from the underlying dynamics, in ${\cal N}=2$, non--linear supersymmetry is imposed by hand through a non--trivial deformation that cannot be obtained by an expectation value of the auxiliary fields. 

We then studied interactions between the goldstino and matter multiplets of ${\cal N}=2$ supersymmetry (vectors and single--tensors that have off--shell descriptions), as well as generalisations of the nilpotent constraints describing incomplete matter multiplets. The interactions are of the Chern--Simons type and describe a super--Brout--Englert--Higgs phenomenon without gravity where the goldstino is absorbed into a massive ${\cal N}=1$ vector multiplet. The constraints describe, in the case of a goldstino in a Maxwell multiplet, either incomplete ${\cal N}=2$ vector multiplets containing only a ${\cal N}=1$ vector, or incomplete (``long'') ${\cal N}=2$ single--tensors containing a ${\cal N}=1$ chiral multiplet. Similarly, in the case of a goldstino in a linear multiplet, the constraints describe either incomplete single--tensors containing a ${\cal N}=1$ linear multiplet, or (``long'') Maxwell containing a ${\cal N}=1$ chiral multiplet. We were not able to find constraints on incomplete ${\cal N}=2$ matter multiplets that do the opposite, keeping the ${\cal N}=1$ linear component of a single--tensor in the first case, or the ${\cal N}=1$  vector component of the Maxwell multiplet in the latter case.

It would be interesting to study the interactions of the Goldstone degrees of freedom of a massive spin--3/2 multiplet consisting of an ${\cal N}=1$ vector and an ${\cal N}=1$ linear multiplet. It is not clear whether our results are sufficient to provide a description of such a system. Another open but related question is the coupling to supergravity realising partial breaking of ${\cal N}=2$ supersymmetry and its rigid limit.

\section{Acknowledgements}

J.-P. D. wishes to thank the LPTHE at UPMC, Paris and CNRS for hospitality and 
support. The work of J.-P.~D. has been supported by the Germaine de Sta\"el franco-swiss bilateral program (project
no.~2015--17). C.M. would like to thank the Albert Einstein Center for Fundamental Physics of the Institute for Theoretical Physics of the University of Bern for very warm hospitality and for financially supporting her stay there.

\renewcommand{\thesection}{\Alph{section}}
\setcounter{section}{0}
\renewcommand{\theequation}{\thesection.\arabic{equation}}


\section{Conventions and some useful identities} \label{AppA}
\setcounter{equation}{0}

The notation $[\ldots]$ in (\ref{part1}) is used for antisymmetrization with weight one. Specifically, 
$$
\partial_{[\mu}B_{\nu\rho]} = {1\over6} \, \partial_\mu \, B_{\nu\rho} \pm \makebox{5 permutations}\,.
$$

The supersymmetric derivatives $D_\alpha$ and $\ov D_\dalpha$ are the usual $N=1$ expressions verifying
$\{ D_\alpha , \ov D_\dalpha \} = -2i (\sigma^\mu)_{\alpha\dalpha}\partial_\mu$:
\beq
D_\alpha = {\partial\over\partial\theta^\alpha} - i(\sigma^\mu\ov\theta)_\alpha\,\partial_\mu, 
\qquad\qquad
\ov D_\dalpha= {\partial\over\partial\ov\theta^\dalpha} - i(\theta\sigma^\mu)_\dalpha\,\partial_\mu.
\eeq
As a consequence,
\beq
[ D_\alpha,\ov{DD} ] = -4i(\sigma^\mu\ov D)_\alpha\partial_\mu \,, \qquad\qquad
[ \ov D_\dalpha,DD ] = 4i(D\sigma^\mu)_\dalpha\partial_\mu\,.
\eeq
In $\widetilde D_\alpha$ and $\ov{\widetilde D}_\dalpha$, $\theta_\alpha$, $\ov\theta_\dalpha$ are 
replaced by $\widetilde\theta_\alpha$, $\ov{\widetilde\theta}_\dalpha$. Note also that $(\ov D_\dalpha)^\dag = - D_\alpha$.

The Maxwell field-strength chiral superfields are defined as
\beq
W_\alpha = -{1\over4}\,\ov{DD}D_\alpha\,V ,
\qquad\qquad
\ov W_\dalpha = -{1\over4}\,DD\ov D_\dalpha\,V,
\qquad\qquad
\ov W_\dalpha = - (W_\alpha)^*
\eeq
where $V$ is a real superfield. In addition,
\beq
\begin{array}{rclrcl}
2i\, \partial_\mu\chi\sigma^\mu\ov{\widetilde\theta} &=& \ov{\widetilde\theta D} \, D\chi ,
\qquad&\qquad \displaystyle
\ov{DD} \, \ov{\widetilde\theta \chi} &=& - 2\,\ov{\widetilde\theta D} \, \ov{D\chi},
\crbig
2\, \widetilde\theta\sigma^\mu\partial_\mu \ov\omega &=& i\, \widetilde\theta^\alpha
\ov D_\dalpha D_\alpha \ov\omega^\dalpha,
\qquad&\qquad
\ov{DD} \, \widetilde\theta \omega 
&=& \widetilde\theta^\alpha\ov D_\dalpha \ov D^\dalpha \omega_\alpha\,,
\end{array}
\eeq
where $\chi_\alpha$ (left--handed) and $\ov\omega_\dalpha$ (right--handed) are ${\cal N}=1$ chiral spinor superfields, $\ov D_\dalpha\chi_\beta = \ov D_\dalpha\ov\omega_\dbeta = 0$, and
\beq
{1\over16}\, \ov{DD} \, DD \, Y  = - \Box Y.
\eeq
where $Y$ is a chiral superfield, $\ov D_\dalpha\,Y=0$. Other useful identities are
$$ 
\ov\eta_\dalpha\,\ov{DD} = -2\,\ov D_\dalpha \, \ov{\eta D} \,, \qquad\qquad 
\ov{DD}\,D_\alpha\mathbb{L} = -2\,\ov D_\dalpha
D_\alpha\ov D^\dalpha\mathbb{L} = 4i (\sigma^\mu\ov D)_\alpha \partial_\mu\mathbb{L}\,,
$$
where $\mathbb{L}$ is a complex linear superfield.

\section{More on the Maxwell supermultiplet} \label{AppB}
\setcounter{equation}{0}

The usual construction of the ${\cal N}=2$ Maxwell multiplet starts with two real ${\cal N}=1$ superfields
$V_1$ and $V_2$ with second supersymmetry variations
\beq
\label{B1}
\delta^* V_1 = -{i\over\sqrt2} (\eta D + \ov{\eta D}) V_2,
\qquad\qquad
\delta^* V_2 = \sqrt2i (\eta D + \ov{\eta D}) V_1.
\eeq
The parameters of the $U(1)$ gauge variations are in a single--tensor ${\cal N}=2$ multiplet:
\beq
\label{B2}
\delta_{gauge} V_1 = \Lambda_\ell \,,
\qquad\qquad
\delta_{gauge} V_2 = \Lambda_c + \ov \Lambda_c \,,
\eeq
with $\Lambda_\ell$ real linear and $\Lambda_c$ chiral:
$\Lambda_\ell = \ov\Lambda_\ell$, $\ov{DD}\,\Lambda_\ell=0$, $\ov D_\dalpha \Lambda_c=0$.
Under the second supersymmetry,
\beq
\label{B3}
\delta^*\Lambda_\ell =- {i\over\sqrt2} (\eta D \Lambda_c + \ov{\eta D} \ov\Lambda_c),
\qquad\quad
\delta^*\Lambda_c = \sqrt2i \, \ov{\eta D}\Lambda_\ell,
\qquad\quad
\delta^*\ov\Lambda_c = \sqrt2i \, \eta D \ov\Lambda_\ell.
\eeq
The gauge field is the $\theta\sigma^\mu\ov\theta$ component of $V_2$.
The ${\cal N}=2$ multiplet containing the field strength $F_{\mu\nu}$ uses the chiral superfields
\beq
\label{B4}
X = {1\over2}\,\ov{DD}\, V_1, 
\qquad\quad
W_\alpha = -{1\over 4}\ov{DD}D_\alpha V_2
\qquad\quad
\ov W_\dalpha = -{1\over 4} DD \ov D_\dalpha V_2.
\eeq
Variations (\ref{B1}) imply:
\beq
\label{B5}
\begin{array}{l}
\delta^*X = \sqrt 2 \, i \, \eta^\alpha W_\alpha ,
\qquad\qquad\qquad\qquad
\delta^*\ov X = \sqrt 2 \, i \, \ov\eta_\dalpha \ov W^\dalpha ,
\crbig
\delta^* W_\alpha =  \sqrt 2 \, i \left[ \frac{1}{4}\eta_\alpha \ov{DD}\,\ov X 
+ i (\sigma^\mu\ov\eta)_\alpha \, \partial_\mu X \right] ,
\crbig
\delta^* \ov W_\dalpha = \sqrt 2 \, i \, \left[ \frac{1}{4} \ov\eta_\dalpha {DD}\, X 
- i (\eta\sigma^\mu)_\dalpha \, \partial_\mu \ov X \right].
\end{array}
\eeq
These are the second supersymmetry variations of the components of the ``short"
chiral--chiral superfield (\ref{Max1}):
$$
{\cal W} (y,\theta,\widetilde\theta)
= X + \sqrt2i\,\widetilde\theta W - {1\over4}\widetilde\theta\widetilde\theta \,\ov{DD}\,\ov X .
$$

To go to the ``long" Maxwell multiplet, one introduces the complex linear $\mathbb{L}$ with eq.~(\ref{CA8}),
\beq
\label{B7}
V_2 = 2 (\mathbb{L} + \ov\mathbb{L}), \qquad\qquad \delta_{gauge} \,\mathbb{L} = {1\over2} \Lambda_c\,,
\eeq
and variations (\ref{B1}) suggest to write
\beq
\label{B6} 
\delta^*\mathbb{L} = {i\over\sqrt2} \, \ov{\eta D} \, V_1,
\qquad\qquad
\delta^*\ov\mathbb{L} = {i\over\sqrt2} \, \eta D \, V_1,
\eeq
which verifies the linearity conditions $\ov{DD} \, \mathbb{L} = DD \, \ov\mathbb{L} = 0$.
However, $\mathbb{L}$ ($12_B+12_F$) and $V_1$ ($8_B+8_F$) do not form an off--shell representation
of ${\cal N}=2$: the algebra does not properly close\,\footnote{See below.} 
and the number of off--shell fields is not a multiple of $8_B+8_F$.

To find the complete multiplet, we rely upon the chiral--antichiral superfield written in its two forms
(\ref{CA1}) and (\ref{CA5}):
\beq
\label{B8}
\begin{array}{rcl}
\widehat{\cal W} &=& U + \sqrt2\, \ov{\widetilde\theta \, \Omega} - \ov{\widetilde\theta\widetilde\theta}\,\Bigl[
{i\over2}X +{1\over4}\ov{DD}\,\ov U \Bigr] ,
\crbig
\widehat{\cal W} &=& U + \sqrt2\, \ov{\widetilde\theta \, D} \, \mathbb{L} - \ov{\widetilde\theta\widetilde\theta}\,\Bigl[
{i\over2}X +{1\over4}\ov{DD}\,\ov U \Bigr].
\end{array}
\eeq
Since the first expression is a chiral--antichiral superfield with $16_B+16_F$ components\footnote{
$U$ and $X$ have $4_B+4_F$ components each, $\Omega_\alpha$ includes $8_B+8_F$ fields.},
the second supersymmetry variations
\beq
\label{B9}
\begin{array}{rcl}
\delta^* U &=& \sqrt2\,\ov{\eta\Omega}, 
\crbig
\delta^*\ov\Omega_\dalpha &=& \displaystyle - {i\over\sqrt2}\Bigl[ X\, \ov\eta_\dalpha 
+ i\,\ov D_\dalpha (\eta DU + \ov{\eta DU}) \Bigr],
\crbig
\delta^* X &=& \displaystyle 2\sqrt2i \Bigl[ {1\over4}\ov{DD}\ov{\eta\Omega} 
- i\,\eta\sigma^\mu\partial_\mu\ov\Omega \Bigr] 
\end{array}
\eeq
give an off--shell representation of ${\cal N}=2$ supersymmetry. 

In the second expression (\ref{B8}), $\ov\Omega_\dalpha$ has been replaced by $\mathbb{L}$, introducing
$4_B+4_F$ supplementary components which are actually invisible in $\widehat{\cal W}$: the gauge
variation (\ref{B7}) leaves $\widehat{\cal W}$ invariant. In addition, the variation 
$\delta^*\Omega_\dalpha$ cannot be written as
$\ov D_\dalpha \delta^*\mathbb{L}$ without a supplementary condition on the chiral $X$. This is where
$$
X={1\over2}\ov{DD} V_1
$$ 
helps by firstly adding $4_B+4_F$ fields to reach $24_B+24_F$ with $U$ and $\mathbb{L}$ and
secondly by turning the second supersymmetry variations (\ref{B9}) into
\beq
\label{B10}
\begin{array}{rcl}
\delta^* U &=& \sqrt2 \, \ov{ \eta D}\,\mathbb{L} \, , \qquad\qquad\qquad
\delta^* \ov U \,\,=\,\, -\sqrt2 \, \eta D\,\ov\mathbb{L} \, ,
\crbig
\delta^* \mathbb{L} &=& \displaystyle {i\over\sqrt2} \, \ov{\eta D}\, V_1 + {1\over\sqrt2}\, (\eta D \, U
+ \ov {\eta D \, U}) \, ,
\crbig
\delta^* \ov\mathbb{L} &=& \displaystyle {i\over\sqrt2} \, \eta D \, V_1 - {1\over\sqrt2}\, (\eta D \, U
+ \ov {\eta D \, U}) \, ,
\crbig
\delta^* V_1 &=& \displaystyle -{i\over\sqrt2} \, (\eta D + \ov{\eta D}) \, 2(\mathbb{L} + \ov\mathbb{L}) 
\,\,=\,\, -{i\over\sqrt2} \, (\eta D + \ov{\eta D}) \, V_2 
\end{array}
\eeq
which represents ${\cal N}=2$ supersymmetry off--shell.\footnote{Verifying explicitly
the closure of the algebra is relatively easy.}
This is the long representation of the Maxwell ${\cal N}=2$ supermultiplet with ${\cal N}=1$ superfield content $U$, $\mathbb{L}$ and $V_1$ for a total of $24_B+24_F$ fields.
Since
\beq
\label{B11}
\delta^* \, V_2 = \sqrt2i\, (\eta D+\ov{\eta D})V_1\,,
\eeq
the $16_B+16_F$ multiplet with superfields $V_1$ and $V_2$ is included in the long 
representation. 

The long multiplet has two gauge variations generated by two independent single--tensor multiplets
with superfields $(\Lambda_\ell,\Lambda_c)$ and $(\Sigma_\ell,\Sigma_c)$ respectively.
The gauge variations are
\beq
\label{B12}
\begin{array}{rcllrcll}
\delta_{gauge}\, U &=& 0 & + \Sigma_c \,, &&&&
\crbig
\delta_{gauge}\, \mathbb{L} &=& {1\over2}\Lambda_c & + i\Sigma_\ell \,,
\qquad&\qquad
\delta_{gauge}\, \ov\mathbb{L} &=& {1\over2}\ov\Lambda_c & -i\Sigma_\ell \,,
\crbig
\delta_{gauge}\, V_1 &=& \Lambda_\ell & +0 \,,
\qquad&\qquad
\delta_{gauge}\, V_2 &=& \Lambda_c+\ov\Lambda_c & +0\, .
\end{array}
\eeq
Standard Maxwell gauge transformations (\ref{B2}) are generated by $(\Lambda_\ell,\Lambda_c)$.
They leave invariant $U$, $\ov D_\dalpha \, \mathbb{L}$, $W_\alpha$ and $X$ and then also the
${\cal N}=2$ superfields ${\cal W}$ and $\widehat{\cal W}$. 

The gauge transformation generated by $(\Sigma_\ell,\Sigma_c)$ acts on $\widehat{\cal W}$ 
according to
\beq
\label{B13}
\delta_{gauge} \, \widehat{\cal W} 
= \Sigma_c + \sqrt2i\, \ov{\widetilde\theta D} \Sigma_\ell 
-\ov{\widetilde\theta\widetilde\theta}\, {1\over4}\,\ov{DD}\,\ov\Sigma_c \equiv {\cal S}.
\eeq
which is a short chiral--antichiral multiplet similar to eq.~(\ref{longM0}). This is the gauge transformation 
already discussed in paragraph \ref{sec51}, which leaves $V_1$, $V_2$, $W_\alpha$, $X$ and then 
${\cal W}$ invariant.
There is a (${\cal N}=1$) gauge in which $U=0$. In this gauge, however,
\beq
\label{B14}
\begin{array}{rcl}
[ \delta_1^* , \delta_2^* ] \, \mathbb{L} &=& 2i \,(\eta_1\sigma^\mu\ov\eta_2 - \eta_2\sigma^\mu\ov\eta_1) \, \partial_\mu\mathbb{L}
-i \Lambda_\ell
\crbig
\Lambda_\ell &=& i (\ov{\eta_2D}\,\eta_1D - \ov{\eta_1D}\,\eta_2D ) \ov\mathbb{L}
- i (\eta_2D\,\ov{\eta_1D} - \eta_1D\,\ov{\eta_2D}) \mathbb{L}\,.
\end{array}
\eeq
Since $\Lambda_\ell$ is a real linear superfield, the algebra closes up to a gauge transformation of $\mathbb{L}$
and the multiplet is not complete without $U$.\footnote{In this gauge, variations (\ref{B6}) hold.}

The two sets of gauge variations (\ref{B12}) remove $16_B+16_F$ components in the long supermultiplet,
to obtain the $8_B+8_F$ physically relevant components of the super-Maxwell theory:
the gauge field $-4\Re\mathbb{V}_\mu$ ($3_B$), the (auxiliary) 
longitudinal vector $D=-4\,\partial^\mu\Im\mathbb{V}_\mu$ ($1_B$), the two complex scalars in $X$ ($4_B$)
and two Majorana gauginos ($8_F$).


\section{More on \boldmath{$\Im \mathbb{V}_\mu$}} \label{AppC}

In the construction of the long Maxwell ${\cal N}=2$ superfield, the abelian gauge field is not, as is usually the case,
a component of a real superfield $V$, but it appears in the expansion of a complex linear superfield $\mathbb{L}$,
with the relation $V=2(\mathbb{L} + \ov{\mathbb{L}})$. As a consequence, the auxiliary scalar field $D$ in
the expansion of $V$ is replaced by the divergence of a vector field. Comparing expansion (\ref{CA4}) of
$\mathbb{L}$ with
$$
V = \theta\sigma^\mu\ov\theta \, A_\mu + {1\over2} \theta\theta\ov{\theta\theta}\, D + \ldots
$$
one finds $A_\mu = -4\Re \mathbb{V}_\mu$ and $D= -4 \, \partial^\mu \Im \mathbb{V}_\mu$. 
In the version of super-Maxwell theory\,\footnote{This Appendix applies to ${\cal N}=1$ and ${\cal N}=2$ 
super-Maxwell theories.} with the auxiliary scalar $D$, its lagrangian is quadratic in $D$:
\beq
\label{appc1}
{\cal L}_D = {1\over2} A\,D^2 + {1\over2} (B+\xi)D,  \qquad\qquad A > 0,
\eeq
where $A$ and $B$ are functions of other scalar fields\,\footnote{They do not depend on derivatives 
of fields. These scalar fields are collectively denoted by $z$.} and the constant $\xi$ is the FI term. 
In particular, $A$ would be the gauge kinetic metric in super-Maxwell theory (hence the positivity condition).
To integrate over $D$, it is legitimate to solve the field equation $2AD + B+ \xi =0$ and substitute the result 
into ${\cal L}_D$ to obtain the scalar potential
\beq
\label{appc2}
{\cal L}_D = - { (B+\xi)^2 \over 8A} = - {\cal V}.
\eeq
This theory does not have any symmetry and the (supersymmetric) ground state is at $\langle B \rangle = -\xi$.
The contribution of ${\cal L}_D$ to the field equations of the scalars appearing as variables of $A$ and $B$
is of course given by
\beq
\label{appc2a}
\partial_z{\cal L}_D = - \partial_z {(B+\xi)^2 \over 8A} = -\partial_z {\cal V}.
\eeq

The replacement $V=2(\mathbb{L}+\ov\mathbb{L})$ leads to $D=\partial^\mu V_\mu$ with
$V_\mu=-4\, \Im \mathbb{V}_\mu$
and then to a quadratic lagrangian for the divergence of a vector field,
\beq
\label{appc3}
{\cal L} = {1\over2} A (\partial^\mu V_\mu)^2 + {1\over2} (B+\xi)\,\partial^\mu V_\mu , \qquad\qquad A > 0 ,
\eeq
instead of expression (\ref{appc1}). Now, the FI term is a derivative which does not contribute
to the dynamical equations and the field equation for $V_\mu$ is
\beq
\label{appc4}
\partial_\mu [ 2A\,\partial^\nu V_\nu + B] = 0 .
\eeq
Its solution 
\beq
\label{appc5}
\partial^\nu V_\nu = - {B+c\over 2A}
\eeq
involves an integration constant $c$ which replaces the FI coefficient $\xi$. The more subtle point is
the procedure to obtain the lagrangian after the integration of $\partial^\mu V_\mu$, since the right-hand side
of the solution is not a derivative of off-shell fields. 

This situation is not new in the literature. Redefine 
\beq
\label{appc9}
V_\mu = {1\over6} \,\epsilon_{\mu\nu\rho\sigma} A^{\nu\rho\sigma},
\qquad\qquad
F_{\mu\nu\rho\sigma} = 4\,\partial_{[\mu}A_{\nu\rho\sigma]}.
\eeq
Since
\beq
\label{appc10}
\partial^\mu V_\mu = {1\over24}\epsilon^{\mu\nu\rho\sigma}F_{\mu\nu\rho\sigma},
\qquad\qquad
(\partial^\mu V_\mu)^2 = - {1\over24}F^{\mu\nu\rho\sigma}F_{\mu\nu\rho\sigma},
\eeq
the lagrangian (\ref{appc3}) becomes
\beq
\label{appc11}
{\cal L}_F = -{1\over48}A \, F^{\mu\nu\rho\sigma}F_{\mu\nu\rho\sigma}
+ {1\over48} (B+\xi) \, \epsilon^{\mu\nu\rho\sigma}F_{\mu\nu\rho\sigma}.
\eeq
It is part of ${\cal N}=8$ supergravity, with $A=e$, and the introduction of the $\xi$ term
has been studied as a potential source for a cosmological constant \cite{ANT}.
Another example is the massive Schwinger model \cite{Cole}\footnote{As also
explained in ref.~\cite{ANT}.} where the Maxwell lagrangian
\beq
{\cal L} = -{1\over4} F_{\mu\nu}F^{\mu\nu} + {1\over2}\theta\, \epsilon^{\mu\nu}\partial_\mu A_\nu
+ A^\mu j_\mu
\eeq
($j_\mu$ is a conserved fermion current)
does not propagate any field. In the gauge $A_0=0$, 
\beq
{\cal L} = {1\over2} (\partial_0A_1)^2 + \theta\, \partial_0A_1,
\eeq
and the field equation $\partial_0^2 A_1= j_1$ implies the presence of a physically-relevant arbitrary integration 
constant in $F_{01}= \partial_0A_1$, to be identified with the parameter $\theta$.

Returning to our lagrangian (\ref{appc3}) and solution (\ref{appc5}),
if we substitute the solution into the lagrangian, $\partial^\mu V_\mu$
becomes a function of the scalar fields $z$, it is not any longer a derivative and the $\xi$--term would then
become physically relevant and contribute to the field equation of $z$. 
We obtain
\beq
\label{appc5a}
{\cal L} = - {(B+\xi)^2 \over 8A} + {(\xi-c)^2 \over 8A} = - {\cal V} 
\eeq
and the contribution of ${\cal L}$ to the field equations of the scalar fields $z$ is of course
$\partial_z{\cal L} = -\partial_z {\cal V}$. Comparing with expression (\ref{appc2a}), 
equivalence is obtained if we identify the integration constant with the FI coefficient $\xi$,
\beq
c = \xi.
\eeq
except if $A$ is constant (the super-Maxwell theory has then canonical kinetic terms), in which case
the second constant term in the potential is irrelevant. 
With this procedure, both versions of the theory 
depend on a single arbitrary constant $c=\xi$, the FI coefficient of the super-Maxwell theory.

Notice that a derivative term may in general contribute to currents. The canonical energy-momentum 
tensor for a ``lagrangian"
${\cal L}_\xi = \xi\,\partial^\mu V_\mu$ is
\beq
\label{appc7}
T_{\mu\nu} = \xi\,[ \partial_\nu V_\mu - \eta_{\mu\nu} \, \partial^\rho V_\rho ] 
\eeq
which is not zero, conserved ($\partial^\mu T_{\mu\nu}=0$) and an improvement term (so that the total 
energy-momentum is zero, assuming the absence of boundary contributions):
\beq
\label{appc8}
T_{00} = \xi\, \vec\nabla\cdot\vec V, \qquad\qquad T_{0i} = \xi\, \partial_i V_0.
\eeq


\end{document}